\documentclass[12pt]{article}

\usepackage{latexsym}
\usepackage{amssymb,amsfonts,amsmath}
\usepackage{graphicx} 
\usepackage{indentfirst}
\usepackage{bbm}
\usepackage{amssymb}
\usepackage{verbatim}
\usepackage{amsmath, amsthm,amssymb}
\usepackage{mathrsfs}
\usepackage{hyperref}
\usepackage{amsfonts}
\usepackage{dsfont}
\usepackage{cite}
\usepackage{xcolor}
\usepackage[multiple]{footmisc}
\parskip=\medskipamount

\newcommand {\cA}{{\cal A}}
\newcommand {\cB}{{\cal B}}

\newcommand {\cD}{{\cal D}}
\newcommand {\cE}{{\cal E}}

\newcommand {\cH}{{\cal H}}
\newcommand {\cI}{{\cal I}}
\newcommand {\cJ}{{\cal J}}

\newcommand {\cL}{{\cal L}}

\newcommand {\cN}{{\cal N}}
\newcommand {\cO}{{\cal O}}

\newcommand {\cQ}{{\cal Q}}

\newcommand {\cT}{{\cal T}}

\newcommand {\cV}{{\cal V}}
\newcommand {\cW}{{\cal W}}

\newcommand {\cZ}{{\cal Z}}
\newcommand{\bD}{{\bf D}}

\newcommand{\bG}{{\bf G}}

\newcommand{\bP}{{\bf P}}

\newcommand{\bX}{{\bf X}}

\newcommand{\bZ}{{\bf Z}}
\def\a{\alpha}
\def\b{\beta}

\def\d{\delta}
\def\e{\epsilon}

\def\g{\gamma}
\def\G{\Gamma}

\def\j{\psi}

\def\l{\lambda}

\def\o{\omega}
\def\p{\pi}
\def\q{\theta}
\def\r{\rho}
\def\s{\sigma}
\def\t{\tau}

\def\x{\xi}
\def\z{\zeta}
\def\D{\Delta}
\def\F{\Phi}
\def\J{\Psi}
\def\L{\Lambda}
\def\O{\Omega}

\def\Q{\Theta}

\def\U{\Upsilon}

\newcommand{\ad}{{\dot{\alpha}}}                           %new
\newcommand{\bd}{{\dot{\beta}}}                            %new
\newcommand{\ve}{\varepsilon}                            %new
\newcommand{\ab}{{\a\b}}

\newcommand{\pa}{\partial}                           %new
\newcommand{\hf}{\frac12}
\newcommand{\abar}{\bar{a}}
\newcommand{\bbar}{\bar{b}}
\newcommand{\cbar}{\bar{c}}
\newcommand{\be}{\begin{equation}}
\newcommand{\ee}{\end{equation}}
\newcommand{\bea}{\begin{eqnarray}}
\newcommand{\eea}{\end{eqnarray}}
\newcommand{\non}{\nonumber}

\numberwithin{equation}{section}

\begin{document}
%%%%%%%%%%%%%%%%%%%%%%%%%
%%%%%%%%%%%%%%%%%%%%%%%%

\begin{titlepage}
\thispagestyle{empty}

\begin{flushright}
LMU-TPW-99-14\\
hep-th/9907107 \\
July, 1999
\end{flushright}

\vspace{1cm}
\begin{center}
{\Large\bf  Correlation Functions of Conserved Currents \\ 
in \mbox{$\cN$} = 2 Superconformal Theory }\\
\end{center}
%\vspace{3mm}

\begin{center} 
{\large Sergei M. Kuzenko\footnote{E-mail: 
sergei@theorie.physik.uni-muenchen.de} 
and Stefan Theisen\footnote{E-mail: 
theisen@theorie.physik.uni-muenchen.de}  }\\
\footnotesize{
{\it Sektion Physik, Universit\"at M\"unchen\\
Theresienstr. 37, D-80333 M\"unchen, Germany} 
} \\
\end{center}
\vspace{1cm}

\begin{abstract}

Using a manifestly supersymmetric formalism, we determine the general 
structure of two- and three- point functions of the supercurrent
and the flavour current of $\cN = 2$ superconformal field theories. 
We also express them in terms of $\cN = 1$ superfields and compare 
to the generic $\cN = 1$ correlation functions. A general discussion of the 
$\cN=2$ supercurrent superfield and the multiplet of anomalies
and their definition as derivatives with respect to the 
supergravity prepotentials is also included. 

\end{abstract}
\vfill
\end{titlepage}

\newpage
\setcounter{page}{1}

\section{Introduction}
\setcounter{footnote}{0}

Superconformal field theories in various dimensions have been intensively
studied for many years. The conjecture of Maldacena \cite{maldacena}, 
which in its
simplest form relates $\cN=4$
super-Yang-Mills theory in four dimensional 
Minkowski space to $\cN=8$ supergravity in five dimensional 
anti-de-Sitter space has led to a renewed interest in
superconformal field theories in diverse dimensions with maximal 
and less than maximal supersymmetry. 
Here we will be interested in $\cN=2$ {\sl generic} superconformally 
invariant theories. Particular examples can be realized as world-volume 
theories on D3 branes in the presence of D7 branes
\cite{D3D7}. 
These theories
have also been studied in the context of the Maldacena conjecture
\cite{D3D7Mal}. 
A more general interest in $\cN=2$ supersymmetric theories, not 
necessarily conformally invariant, arises within the context of 
Seiberg-Witten theory and its string/M-theory realization.
For reviews, see e.g. \cite{aharonyetal,AGH,GK}. 

A general efficient formalism to analyse correlation functions
of quasi-primary fields 
has been developed since the early days of 
conformal field theory. 
Some important recent contributions have been provided by 
Osborn and collaborators. We refer to their papers: 
to ref. \cite{op,eo} for the non-supersymmetric case in an arbitrary 
number of dimensions. 
A complete analysis of the $\cN=1$ supersymmetric case in $d=4$ 
was presented in \cite{osborn} (see also \cite{park1}). 
In ref. \cite{park2} Park constructed the 
building blocks of correlators of quasi-primary fields for 
arbitrary $\cN$ in four dimensions and for $(p,0)$ superconformal 
symmetry in $d=6$. The formalism is powerful for applications
whenever there exist off-shell superfield formulations for 
superconformal theories, and such formulations are known in four
dimensions for $\cN=1,~2,~3$.

In this paper we are going to analyse correlation functions of conserved 
currents in $\cN=2$, $d=4$ superconformal field theory in a manifestly 
$\cN=2$ supersymmetric language. To this end we review in sect. 2 
the formalism 
of Osborn and Park, specializing to the case of $\cN=2$. 
In sect. 3 we 
apply this to the computation of various two- and three-point correlation
functions, involving the $\cN=2$ supercurrent ${\cal J}$
and flavour currents ${\cal L}^{ij}$.
The three-point function of the supercurrent 
is shown to be the sum of two linearly 
independent superconformal structures whose coefficients are related 
to the anomaly coefficients, denoted by $a$ and $c$ in \cite{Anselmietal}.
Whereas for
$\cN=1$ there exist two independent structures for the three-point 
function of the flavour current,  
there is only one for 
$\cN=2$. This is a consequence of the fact that $\cN=2$ 
theories are non-chiral.
We also analyse mixed three-point functions and, in particular, show 
that the three-point function $\langle \cJ\, \cJ\,{\cal L}^{ij} \rangle$
vanishes, as a consequence of $\cN=2$ superconformal symmetry.  
In sect. 4 we describe the reduction of our results to $\cN=1$ 
superfields. The main body of the paper ends with a brief discussion. 
We have included a few technical appendices to make the paper self-contained. 
In App. A we review the 
Weyl and the minimal $\cN=2$ supergravity multiplets in harmonic superspace
and present a new parametrisation of the supergravity prepotential
(it was sketched already in part by Siegel \cite{siegel})
which is most convenient for any consideration involving
the supercurrent and the multiplet of anomalies. 
In App. B we describe the procedure to generate the supercurrent and the 
multiplet of anomalies as functional derivatives with respect to 
supergravity prepotentials. In App. C we compute the supercurrent and the 
multiplet of anomalies for general renormalizable 
$\cN=2$ super-Yang-Mills models. 

The multiplets of currents and anomalies for $\cN=2$ extended supersymmetry 
in four space-time dimensions were introduced by Sohnius \cite{sohnius}
twenty years ago. He considered the simplest $\cN=2$ supersymmetric model --
the hypermultiplet with $8+8$ off-shell degrees of freedom \cite{f-s}, and showed
that the energy-momentum tensor $\Theta_{mn}$
belongs to a supermultiplet 
(called, by analogy with $\cN=1$ SUSY \cite{fz}, 
the $\cN=2$ supercurrent) which
(i) in addition, contains  
the SU(2) $R$-current $j^{(ij)}_m$, the axial current
$j^{(R)}_m$, the $\cN=2$ supersymmetry currents
$j^i_{m \hat{\a}}$, where $\hat{\a} = \a , \ad $, 
the central charge current $c_m$
as well as some auxiliary components of lower dimension;
(ii) is described by a real scalar superfield $\cJ(z)$ of mass dimension 2.
The central charge current is also part of the multiplet of anomalies
which contains in addition $\Theta^m{}_m$, $\pa^m j^{(R)}_m$ and
$(\g^m j^i_m)_{\hat{\a}}$
along with an auxiliary triplet. 
The multiplet of anomalies
is described by a real isotriplet superfield $\cT^{(ij)}(z)$, 
$\overline{ \cT^{ij} } = \cT_{ij}$, which is subject to the constraint
\be
D^{(i}_\a \,\cT^{jk)} = {\bar D}^{(i}_\ad \,\cT^{jk)} = 0
\label{trace}
\ee
where $D_A= (\pa_a \,, \, D^i_\a\, , \, {\bar D}^\ad_i )$ are 
the $N=2$ supersymmetric
covariant derivatives, 
$i = \underline{1},\underline{2}$.
Both $\cJ$ and $\cT^{ij}$ turn out
to be invariant with respect to the 
central charge transformations.
The supercurrent conservation law reads
\be
\frac{1}{4}D^{ij} \cJ + {\rm i} \,\cT^{ij} = 0 \qquad
\Longleftrightarrow \qquad
\frac{1}{4} {\bar D}^{ij} \cJ - {\rm i}\, \cT^{ij} = 0
\label{sccl}
\ee 
where $D^{ij} = D^{\a ( i} D^{j)}_\a$, 
${\bar D}^{ij}
= {\bar D}^{(i}_\ad {\bar D}^{j) \ad }$. 
The constraint (\ref{trace}) means  
that $\cT^{ij}$ is a so-called
$\cN=2$ linear multiplet. Such a multiplet
contains a conserved vector and the reality 
condition for $\cT^{ij}$ 
is equivalent to the absence of the second 
(fundamental) central charge
(which is the case for all
$\cN=2$ irreducible supermultiplets). 

A nice feature of the $\cN=2$ multiplet of 
anomalies is that its supersymmetric
structure is completely analogous to that of 
a $\cN=2$ superfield containing 
a conserved flavour current 
of a $\cN=2$ supersymmetric field theory.
Such a flavour current superfield 
$\cL^{(ij)}(z)$,
$\overline{\cL^{ij}}= \cL_{ij} $
satisfies the same constraint,
\be
D^{(i}_\a \,\cL^{jk)} = {\bar D}^{(i}_\ad \,\cL^{jk)} = 0~.
\label{fcs1}
\ee
The similarity is not accidental. The point is 
that $\cL^{ij}$ is generated by coupling matter
hypermultiplets to a gauge vector supermultiplet.
On the other hand, the source for $\cT^{ij}$ 
is a vector multiplet which gauges the central charge 
and belongs to the $\cN=2$ supergravity multiplet.

The structure of the $\cN=2$ supercurrent has been used 
by Sohnius and West  \cite{sw} in their proof of
finiteness of the $\cN=4$  SYM theory which was
based on anomaly considerations.
It is worth pointing out that the 
supercurrent conservation law
in quantum $\cN=2$ super Yang-Mills 
theories \cite{west} (see also \cite{marc})
\be
D^{ij} \cJ = -\frac{1}{3} 
\frac{\b (g)}{g} {\bar D}^{ij} {\bar W}^2
\ee
can be brought to the form (\ref{sccl}) by a 
finite local shift of $\cJ$, resulting in
\be
D^{ij} \cJ = \frac{1}{3}\,
\frac{\b (g)}{g} \left(
D^{ij} W^2 -
{\bar D}^{ij} {\bar W}^2 \right) \;.
\ee
Here $W$ is the $\cN=2$ Yang-Mills field strength,
and $\b (g)$ is the beta-function of the 
gauge coupling constant.

Another consequence of the structure of the
$\cN=2$ supercurrent follows
from the fact that $\cJ$ presents itself
the multiplet of superconformal 
currents. Then, Noether's procedure tells
us that  $\cN=2$ conformal supergravity should be 
described by a real scalar prepotential $\bG(z)$ \cite{hst,rt}
to which the matter supercurrent is coupled.
In App. A.1 we will show how such a prepotential 
arises in the harmonic superspace 
approach to $\cN=2$ conformal supergravity
\cite{gios1,gios2}. This point requires
some comments. 
Many years ago, Gates and Siegel
\cite{gs} showed that the first minimal $\cN=2$ Poincar\'e
supergravity (in the terminology of the third reference
in \cite{dwsugra}) is described, at the linearized level, 
by a single unconstrained
spinor superfield $\J_{\a i}(z)$.\footnote{The harmonic superspace origin
of this prepotential has been recently revealed by Zupnik \cite{zupnik}.}
Their conclusion is in perfect
agreement with the fact that (i) the corresponding 
superspace differential geometry \cite{gates} contains
two independent strengths - a covariantly chiral symmetric bi-spinor
$W_{\a \b}$ ($\cN=2$ super Weyl tensor) and a spinor  $T_{\a i}$ ; (ii)
the supergravity equation of motion reads
\be
 \frac{\d S_{{\rm sugra}} }{\d \J_{ \a  i} } 
~ \propto ~   T^{\a i} ~=~0\;.
\ee
In  \cite{gs} it was argued 
that $\cN=2$ conformal supergravity should be described by the same 
prepotential $\J_{\a i}$ but with a larger gauge freedom.
This led Gates, Grisaru and Siegel \cite{ggs}
to postulate that the $\cN=2$ supercurrent be a spinor superfield
\be
\cJ^i_\a = \frac{\d S_{{\rm matter}} }{\d \J^\a_i}\;.
\ee
As will be described below, this puzzle can be  
resolved in the harmonic superspace approach to $\cN=2$
supergravity \cite{gios1,gios2}. There, 
the prepotential $\bG$ is part of a larger harmonic multiplet 
 $G(z,u)$ with a huge gauge symmetry. The gauge freedom 
can be fixed  in part either to leave a single 
real unconstrained $\bG  (z)$, the leading component of 
 $G(z,u)$ in its harmonic Fourier expansion, or 
to bring $G(z,u)$ to the form 
\be
G(z,u) = D^{\a i} \J^j_{\a } (z) u^+_i u^{-}_j   ~~+~~ {\rm conjugate} 
\ee
with $\J_{\a i } (z) $ the Gates-Siegel prepotential.
Therefore we have
$\cJ^i_\a = D^i_\a \, \cJ $
for all (renormalizable) $\cN=2$ matter systems.
The details of this discussion are provided in Apps. A and C. 

Manifestly supersymmetric techniques 
to study the quantum dynamics and to compute the 
superconformal anomalies for $\cN=2$
matter systems in a supergravity background are not yet available. 
In $x$--space, there exists an exhaustive
description of general $\cN=2$ supergravity-matter
systems \cite{dwsugra,dwmatter}.
In superspace, there exist elaborated 
differential geometry formalisms
\cite{gates,howe,muller} corresponding
to $\cN=2$ conformal supergravity 
and the three known versions of $\cN=2$
Poincar\'e supergravity.
Moreover,
the unconstrained prepotentials and the gauge
group of $\cN=2$ conformal supergravity 
were found in harmonic superspace \cite{gios1,gios2},
and this analysis was extended to describe different 
versions of $\cN=2$ Poincar\'e supergravity 
\cite{gios2,gio} and most general supersymmetric
sigma models in curved harmonic superspace
\cite{bgio}. What is still missing is the detailed
relationship between the differential superspace geometry 
of $\cN=2$ supergravity \cite{gates,howe,muller} and its
description in terms of the unconstrained prepotentials
given in \cite{gios1,gios2}. Another missing prerequisite
is the definition of the $\cN=2$ supercurrent and 
multiplet of anomalies as the response of the 
$\cN=2$ matter action (in the full nonlinear theory) 
to small disturbances in supergravity prepotentials, 
similar to what is well known in 
$\cN=1$ supersymmetry (see
\cite{bk} for a review)
\be
J_{\a \ad} = \frac{\d S}{\d H^{\a \ad}}~,
\qquad \quad T= \frac{\d S}{\d \varphi}\; ;
\label{n1supercurr}
\ee
here $H^{\a \ad}$ and $\varphi$ are the $\cN=1$ 
gravitational superfield and chiral compensator,
respectively. Such a definition is of primary importance,
since it allows us to compute correlators with supercurrent
insertions simply as functional derivatives of the
renormalised effective action with respect 
to supergravity prepotentials.
In the appendices we will close some of these gaps.
In particular, using the harmonic 
superspace approach to $\cN=2$ supergravity
\cite{gios1,gios2}, which we briefly review,  
we introduce a new parametrisation
of the supergravity prepotentials which allows us
to easily obtain the $\cN=2$ analogue
of (\ref{n1supercurr}).

Before closing this introductory section, 
we would like to comment on the $\cN=1$ 
multiplets contained in $\cJ$ and $\cT^{ij}$ (see also 
\cite{ggs}). For that purpose
we introduce the $\cN=1$ spinor covariant derivatives 
$D_\a \equiv D^{\underline{1}}_\a$, ${\bar D}^\ad \equiv
{\bar D}^\ad_{\underline{1}}$ and define the $\cN=1$ projection 
$U| \equiv U(x,\q^\a_i, {\bar \q}^j_\ad ) 
|_{ \q_{\underline{2}} = 
{\bar \q}^{\underline{2}} = 0}$ of an arbitrary $\cN=2$ superfield
$U$. It follows from (\ref{trace}) and (\ref{sccl}) that 
$\cJ$ is composed of three independent $\cN=1$ 
multiplets
\be
J \equiv \cJ| ={\bar J}~, \qquad  J_\a \equiv
D^{ \underline{2} }_\a \cJ |~, \qquad 
J_{\a \ad} \equiv \hf [D^{ \underline{2} }_\a \, , \, 
{\bar D}_{\ad \underline{2} }] \cJ|
-\frac{1}{6}  [D^{ \underline{1} }_\a \, , \, 
{\bar D}_{\ad \underline{1} }] \cJ| ={\bar J}_{\a \ad}
\label{n=1s-ccomponents}
\ee
while $\cT$ contains two independent $\cN=1$ components
\bea
& T \equiv {\rm i} \cT^{ \underline{22}}|~, \qquad \quad 
{\bar D}_\ad T = 0 \non \\
&L \equiv {\rm i} \cT^{ \underline{12}}| = \bar L~, \qquad \quad 
{\bar D}^2  L = 0
\eea
with $T$ being a chiral superfield, and $L$ a real linear superfield.
It is easy to find the equations for $J$, $J_\a$ and $J_{\a \ad}$:
\bea
&  \frac{1}{4} \,{\bar D}^2 J = T \non \\
&  \frac{1}{4} \,D^\a J_\a = - L~,  \qquad \qquad
{\bar D}^2 J_\a = 0 \non \\ 
& {\bar D}^\ad J_{\a \ad} = \frac{2}{3}\, {\rm D}_\a T \;.
\eea
The latter equation shows that $J_{\a \ad}$ is the $\cN=1$ 
supercurrent and $T$ the corresponding multiplet of anomalies.
The spinor object $J_\a$ contains the second supersymmetry 
current, the central charge current and two of the three 
SU(2) currents, namely those
which correspond to the symmetries belonging to SU(2)$/$U(1).
Finally, the scalar $J$ contains the current corresponding 
to the special combination of the $\cN=2$ U(1) $R$-transformation 
and SU(2) $z$-rotation which leaves $\q_{\underline{1}}$ 
and ${\bar \q}^{\underline{1}}$ invariant.
The central charge current is also contained in $L$, 
which is no accident. In $\cN=1$ supersymmetry,
associated with any internal symmetry is a real linear 
superfield containing the corresponding  conserved current;
$L$ is such a superfield for the central charge.
Similarly, in a superconformal theory ($\cT^{ij} = 0$)
the real scalar $J$ becomes a linear superfield
and, hence, contains a conserved current.  

\section{Superconformal building blocks}
\setcounter{footnote}{0}

\subsection{Superconformal Killing vectors}
In $\cN$--extended global superspace 
${\Bbb R}^{4|4\cN}$ 
parametrised by 
$z^A = (x^a, \q^\a_i, {\bar \q}^i_\ad) $,
infinitesimal superconformal transformations 
\be
z^A \quad \longrightarrow \quad z^A + \x z^A
\ee
are generated by 
superconformal Killing vectors \cite{BPT,bk,west,HH}
\be
\x = {\overline \x} = \x^a (z) \pa_a + \x^\a_i (z)D^i_\a
+ {\bar \x}_\ad^i (z) {\bar D}^\ad_i
\ee   
defined to satisfy 
\be
[\x \;,\; D^i_\a ] \; \propto \; D^j_\b \;.
\ee   
${}$From here one gets
\be
\x^\a_i = -\frac{\rm i}{8} {\bar D}_{\bd i} \x^{\bd \a}\;, \qquad
{\bar D}_{\bd j} \x^\a_i = 0
\label{spinsc}
\ee
while the vector parameters satisfy the master equation
\be
D^i_{(\a} \x_{\b )\bd} = {\bar D}_{i(\ad} \x_{\b \bd )}=0
\label{msc}
\ee
implying, in turn, the conformal Killing equation
\be
\pa_a \x_b + \pa_b \x_a = \hf\, \eta_{ab}\, \pa_c \x^c\;.
\ee 
The general solution of eq. (\ref{msc}) was given 
in \cite{bk} for $\cN=1$ and in \cite{park2} for $\cN >1$.
${}$From eqs. (\ref{spinsc}) and (\ref{msc}) it follows
\be
[\x \;,\; D^i_\a ] = - (D^i_\a \x^\b_j) D^j_\b
= \hat{\o}_\a{}^\b  D^i_\b - \frac{1}{\cN}
\Big( (\cN-2) \s + 2 {\bar \s}  \Big) D^i_\a
-{\rm i} \hat{\L}_j{}^i \; D^j_\a \;.
\ee
Here the parameters of `local' Lorentz $\hat{\o}$ and
scale--chiral $\s$ transformations are
\be
\hat{\o}_{\a \b}(z) = -\frac{1}{\cN}\;D^i_{(\a} \x_{\b)i}\;,
\qquad \s (z) = \frac{1}{\cN (\cN - 4)}
\left( \hf (\cN-2) D^i_\a \x^\a_i - 
{\bar D}^\ad_i {\bar \x}_{\ad}^{ i} \right)
\label{lor,weyl}
\ee
and turn out to be chiral
\be
{\bar D}_{\ad i} \hat{\o}_{\a \b}~=~ 0\;,
\qquad {\bar D}_{\ad i} \s ~=~0\;.
\ee
The parameters $\hat{\L}_j{}^i$ 
\be
\hat{\L}_j{}^i (z) = -\frac{1}{32}\left(
[D^i_\a\;,{\bar D}_{\ad j}] - \frac{1}{\cN}
\d_j{}^i  [D^k_\a\;,{\bar D}_{\ad k}] \right)\x^{\ad \a}~, \qquad
\hat{\L}^\dag =  \hat{\L}~, \qquad  {\rm tr}\; \hat{\L} = 0
\ee
correspond to `local' SU($\cN $) transformations.
One can readily check the identity 
\be
D^k_\a \hat{\L}_j{}^i = 2{\rm i} \left( \d^k_j D^i_\a 
-\frac{1}{\cN} \d^i_j D^k_\a  \right) \s~.
\ee
{}For $\cN=2$ it leads to the analyticity condition
\be
D^{(i}_\a \hat{\L}^{jk)} = {\bar D}^{(i}_\ad \hat{\L}^{jk)}= 0~, 
~~~~~\qquad \qquad 
(\cN =2)~.
\ee

As is seen from (\ref{lor,weyl}), the above formalism
cannot be directly applied to the case $\cN = 4$
which is treated in more detail, e.g., in \cite{park2}.
In what follows, our considerations will be restricted
to $\cN <4$, with special emphasis on the choice $\cN = 2$
later on. 

The superalgebra of $\cN$--extended superconformal Killing vectors
is isomorphic to the superalgebra su$(2,2|\cN)$ spanned by elements
of the form
\be
{\bf g} = \left(
\begin{array}{ccc}
\o_\a{}^\b - \D \d_\a{}^\b  \quad &  -{\rm i} b_{\a \bd} \quad &
2\eta_\a{}^j \\
 -{\rm i} a^{\ad \b} \quad & -{\bar \o}^\ad{}_\bd 
+ {\bar \D}  \d^\ad{}_\bd   \quad &
2{\bar \e}^{\ad j} \\
2\e_i{}^\b \quad & 2{\bar \eta}_{i \bd} \quad & \frac{2}{\cN}({\bar \D} - \D)
\d_i{}^j + {\rm i}\; \L_i{}^j
\end{array}
\right)
\ee
which satisfy the conditions
\be
{\rm str} \;{\bf g} = 0~, \qquad
B{\bf g}^\dag B = - {\bf g}~, \qquad 
B = \left( 
\begin{array}{ccc}
0 \quad & 1 \quad & 0 \\
1 \quad & 0 \quad & 0 \\
0 \quad & 0 \quad & -1 
\end{array}
\right)~.
\ee
Here the matrix elements correspond to a
Lorentz transformation $(\o_\a{}^\b,~{\bar \o}^\ad{}_\bd)$,
translation $a^{\ad \a}$, special conformal transformation
$ b_{\a \ad}$, $Q$--supersymmetry $(\e_i^\a,~ {\bar \e}^{\ad i})$,
$S$--supersymmetry $(\eta_\a^i,~{\bar \eta}_{i \ad})$,
combined scale and chiral transformation $\D$, 
and chiral SU$(\cN)$ transformation $\L_i{}^j$.
They are related to the parameters of the superconformal 
Killing vector as follows
\bea
& \o_\a{}^\b = \hat{\o}_\a{}^\b (z=0)~, \qquad \D =\s (z=0)~,\qquad
\L_i{}^j = \hat{\L}_i{}^j(z=0)~, \non \\
& a^m = \x^m (z=0)~,\qquad \e^\a_i = \x^\a_i (z=0)~,
\eea
and so on. For such a correspondence, $\x \longrightarrow {\bf g}$,
we have
\be
[\x_1\;,\; \x_2] \quad  \longrightarrow \quad -\;[{\bf g}_1\; ,\; {\bf g}_2]~.
\ee 

It is useful to identify Minkowski superspace
as a homogeneous space of the superconformal group SU$(2,2|\cN)$
using the above matrix realization  
\be
\O (z) = \exp \,{\rm i} \left\{ - x^a P_a 
+ \q^\a_i Q^i_\a  + {\bar \q}^i_\ad {\bar Q}^\ad_i \right\} 
= \left(
\begin{array}{rcr}
\d_\a{}^\b \quad &      0  \quad  &      0  \\
-{\rm i} x_+^{\ad \b} \quad& \d^\ad{}_\bd  \quad & 2{\bar \q}^{\ad j} \\
2\q_i{}^\b  \quad & 0 \quad & \d_i{}^j  
\end{array} \right)
\ee
where $x_\mp $ denote ordinary (anti-)chiral bosonic variables
\be
x^a_\pm = x^a \pm {\rm i} \q_i \s^a {\bar \q}^i\;.
\ee
One verifies that 
\bea
& {\bf g}\; \O (z)= \x \O (z) + \O (z)\; {\bf h}(z)~, \non \\
\noalign{\noindent where}
& {\bf h}(z) = \left(
\begin{array}{ccc} 
\hat{\o}_\a{}^\b  - \s \d_\a{}^\b \quad &  -{\rm i} b_{\a \bd} \quad &
2 \hat{\eta}_\a{}^j \\
 0 \quad & - \hat{{\bar \o}}^\ad{}_\bd 
+  {\bar \s}  \d^\ad{}_\bd \quad &
0  \\
0  \quad & 2 \hat{ {\bar \eta}}_{i \bd} \quad & 
\frac{2}{\cN}( {\bar \s} - \s  )
\d_i{}^j + {\rm i}\; \hat{\L}_i{}^j
\end{array}
\right)
\label{stability}
\eea
belongs to the Lie algebra of the 
stability group. Here $\hat{\eta}$ is
\be
\hat{\eta}_\a{}^i (z) = \hf D^i_\a \s (z)~.
\ee
This should be interpreted within the framework of nonlinear realizations.

\subsection{Two-point structures}
Given two points $z_1$ and $z_2$ in superspace,
it is useful to introduce (anti-)chiral combinations
\bea
x^a_{{\bar 1}2} &=& - x^a_{2 {\bar 1}} =  x^a_{1-} - x^a_{2+} 
+2{\rm i} \q_{2 i}\, \s^a \,{\bar \q}_1^i \non \\
\q_{12}&=& \q_1 - \q_2 \qquad \quad {\bar \q}_{12} 
= {\bar \q}_1 - {\bar \q}_2
\eea  
which are invariants of the $Q$--supersymmetry
transformations (the notation `$x_{{\bar 1}2}$' indicates
that $x_{{\bar 1}2}$ is antichiral with respect to $z_1$
and chiral with respect to $z_2$). 
As a consequence of (\ref{stability}), they transform 
semi-covariantly with respect to the superconformal group
\bea
\d x^{\ad \a}_{{\bar 1}2} &=& -\left( \hat{{\bar \o}}^\ad{}_\bd (z_1)   
- \d^\ad{}_\bd \,{\bar \s}(z_1)\right) x^{\bd \a}_{{\bar 1}2}  
- x^{\ad \b}_{{\bar 1}2} \left( \hat{\o}_\b{}^\a (z_2) -
\d_\b{}^\a \,\s (z_2) \right)\non \\
\d \q^\a_{12\,i} &=& {\rm i} \Big( \hat{\L}_i{}^j (z_1)
+\frac{2}{\cN}\left({\bar \s}(z_1) - \s (z_1)\right)
\,\d_i{}^j\Big)
\q^\a_{12\,j} - {\rm i} \,\hat{{\bar \eta}}_{\bd i}(z_1)
x^{\bd \a}_{{\bar 1}2}
\non \\
&& - \q^\b_{12\,i} 
\left( \hat{\o}_\b{}^\a (z_2) -
\d_\b{}^\a \,\s (z_2) \right)
\label{two-point-variation} 
\eea 

{}Following \cite{park2}, it is useful to introduce a 
conformally covariant $\cN \times \cN$ matrix
\footnote{We use the notation adopted  in \cite{wb,bk}.
When the spinor indices are not indicated explicitly,
the following matrix-like conventions are assumed \cite{osborn}:
$\j = (\j^\a)$, $\tilde{\j} = (\j_\a)$,
$\bar{\j} = ( \bar{\j}^\ad)$, 
$\tilde{\bar{\j}} = (\bar{\j}_\ad)$, 
$x = (x_{\a \ad})$, $\tilde{x} =(x^{\ad \a})$;
but $x^2 \equiv x^a x_a = - \hf \,{\rm tr}\, (\tilde{x} x)$,
and hence $\tilde{x}^{-1} = - x / x^2$.
}
\be
u_i{}^j (z_{12}) = \d_i{}^j - 4{\rm i}\;
\frac{ \q_{12\,i}x_{ {\bar 1} 2} {\bar \q}^j_{12} }
{x_{ {\bar 1} 2}{}^2} =
\d_i{}^j + 4{\rm i}\;
\q_{12\,i} \tilde{x}_{ {\bar 1} 2}{}^{-1} {\bar \q}^j_{12}
\ee
with the basic properties
\be
u^\dag (z_{12})~ u (z_{12}) =  1~,\qquad u^{-1} (z_{12}) 
= u (z_{21})~, \qquad \det \; u (z_{12}) = 
\frac{x_{ {\bar 1} 2}{}^2}  {x_{ {\bar 2} 1}{}^2}~.
\label{unitary}
\ee
In accordance with (\ref{two-point-variation}), 
the unimodular unitary matrix 
\be
\hat{u}_i{}^j (z_{12}) = \left( 
\frac{ x_{ {\bar 2} 1}{}^2 }{x_{ {\bar 1} 2}{}^2} \right)^{\frac{1}{\cN} } 
u_i{}^j (z_{12})
\label{unimod1}
\ee
transforms as 
\be
\d \hat{u}_i{}^j (z_{12}) = {\rm i}\,
\hat{\L}_i{}^k (z_1) \hat{u}_k{}^j (z_{12}) -
{\rm i}\,\hat{u}_i{}^k (z_{12})\hat{\L}_k{}^j (z_2) ~.
\ee

\subsection{Three-point structures}
Given three superspace points $z_1,~z_2$ and $z_3$,
one can define superconformally covariant bosonic 
and fermionic variables $\bZ_1,~ \bZ_2$ and $\bZ_3$,
where $\bZ_1 = (\bX_1^a, ~ \Q^{\a i}_1,~ 
{\bar \Q}^\ad _{1\, i})$ are
\cite{osborn,park2}
\bea
\bX_1 & \equiv &\tilde{x}_{1 \bar{2}}{}^{-1} 
\tilde{x}_{\bar{2} 3} \tilde{x}_{3 \bar{1}}{}^{-1} ~, \qquad \quad
\bar{\bX}_1 = \bX^\dag_1 = -\tilde{x}_{1 \bar{3}}{}^{-1}
\tilde{x}_{\bar{3} 2} \tilde{x}_{2 \bar{1}}{}^{-1} \non \\
\tilde{\Q}_1^i & \equiv & {\rm i}\, 
\left( \tilde{x}_{ \bar{2} 1}{}^{-1} {\bar \q}^i_{12} 
- \tilde{x}_{ \bar{3} 1}{}^{-1} {\bar \q}^i_{13} \right)
=\frac{1}{4}\, \tilde{D}^i_{1} \ln \frac{ x_{\bar{2} 1}{}^2 }
{x_{ \bar{3} 1}{}^2 } \non \\
\tilde{ \bar{\Q}}_{1\,i} & \equiv  & {\rm i}\, 
\left( \q_{12\,i} \tilde{x}_{ \bar{1} 2}{}^{-1}
- \q_{13\,i} \tilde{x}_{ \bar{1} 3}{}^{-1} \right)
=\frac{1}{4}\, \tilde{{\bar D}}_{1\,  i} 
\ln \frac{ x_{\bar{1} 2}{}^2 }
{x_{ \bar{1} 3}{}^2 }   
\label{capZ}
\eea
and $\bZ_2,~\bZ_3$ are obtained from here by 
cyclically permuting indices.
These structures possess remarkably simple 
transformation rules under superconformal transformations:
\bea
\d \bX_{1\, \a \ad} & = & 
\left( \hat{\o}_\a{}^\b (z_1) - \d_\a{}^\b \,\s (z_1) \right)
\bX_{1\, \b \ad} + \bX_{1\, \a \bd}
\left( \hat{{\bar \o}}^\bd{}_\ad (z_1)   
- \d^\bd{}_\ad  \,{\bar \s}(z_1)\right) \non \\
\d \Q^i_{1\, \a} & = & 
\hat{\o}_\a{}^\b (z_1) \Q^i_{1\, \b} -
{\rm i} \Q^j_{1\, \a} \hat{\L}_j{}^i (z_1)
-\frac{1}{\cN} \Big( (\cN - 2) \s (z_1) + 
2{\bar \s} (z_1) \Big)\Q^i_{1\, \a} 
\eea
and turn out to be essential building blocks
for correlations functions of quasi-primary 
superfields.

Among important properties of $\bZ$'s are the following:
\bea
&& \bX_1{}^2 = \frac{ x_{\bar{2}3}{}^2 }{ x_{\bar{2}1}{}^2 
x_{\bar{1}3}{}^2 }~,
\qquad \qquad {\bar \bX}_1{}^2 =
\frac{ x_{\bar{3}2}{}^2 }{ x_{\bar{3}1}{}^2 x_{\bar{1}2}{}^2 }\non \\
&& {}\qquad  \bX_{1 \, \a \ad} - {\bar \bX}_{1 \, \a \ad} =
4{\rm i}\, \Q^i_{1\, \a} {\bar \Q}_{1\, \ad i}
\label{bosferm}
\eea
and further relations obtained by cyclic permutation of
labels. The variables $\bZ$'s with different labels are related to each
other:
\bea
& \tilde{x}_{\bar{1} 3} \bX_3 \tilde{x}_{\bar{3} 1} =
- {\bar \bX}_1{}^{-1} ~, \qquad \qquad
\tilde{x}_{\bar{1} 3}{\bar \bX}_3 \tilde{x}_{\bar{3} 1} =
- \bX_1{}^{-1} \non \\
& \tilde{x}_{\bar{1} 3} \tilde{\Q}^i_3 u_i{}^j(z_{31})
= - \bX_1{}^{-1} \tilde{\Q}^j_1~, \qquad
u_i{}^j(z_{13}) \tilde{ {\bar \Q}}_{3\,j} \tilde{x}_{\bar{3} 1}
= \tilde{ {\bar \Q}}_{1\,i} {\bar \bX}_1{}^{-1}~.
\label{difz}
\eea

With the aid of the matrices $u (z_{rs})$, $r,s = 1,2,3$, 
defined in (\ref{unitary}),
one can construct unitary matrices \cite{park2}
\bea
{\bf u} (\bZ_3) &=& u (z_{31}) u (z_{12})u (z_{23})~, \qquad
{\bf u}_i{}^j (\bZ_3) = \d_i{}^j 
- 4{\rm i} \tilde{ \bar{\Q} }_{3\,i} \bX_3{}^{-1} 
\tilde{\Q}^j_3 \non \\
{\bf u}^\dag (\bZ_3) &=& u (z_{32}) u (z_{21})u (z_{13})~, \qquad
{\bf u}^\dag_i{}^j(\bZ_3)=  \d_i{}^j + 4{\rm i}
\tilde{ \bar{\Q} }_{3\,i} {\bar \bX}_3{}^{-1} 
\tilde{\Q}^j_3
\label{greatu}
\eea
transforming at $z_3$ only. Their properties are
\be
{\bf u}^\dag (\bZ_3)= {\bf u}^{-1} (\bZ_3)~,\qquad
\qquad 
\det  {\bf u} (\bZ_3) = \frac{ \bX_3{}^2 }
{ {\bar \bX}_3{}^2 }~.
\ee
It is worth noting that $\det  {\bf u} (\bZ_3)$ is 
a superconformal invariant \cite{park2} and from 
(\ref{bosferm}) one immediately gets
\be
\frac{ \bX_1{}^2 }{ {\bar \bX}_1{}^2 }
=\frac{ \bX_2{}^2 }{ {\bar \bX}_2{}^2 }
=\frac{ \bX_3{}^2 }{ {\bar \bX}_3{}^2 }~.
\ee

\subsection{Specific features of \mbox{$\cN$} = 2 theory}
In the case $\cN = 2$, we have at our disposal 
the SU(2)--invariant tensors $\ve_{ij} = -\ve_{ji} $
and $\ve^{ij} = -\ve^{ji} $, normalized to
$\ve^{\underline{12}}= \ve_{\underline{21}} =1$.
They can be used
to raise and lower isoindices
\be
C^i =  \ve^{ij} C_j ~, \qquad \qquad      
C_i =  \ve_{ij} C^j~.
\ee
Now, the condition of unimodularity of the matrix 
defined in (\ref{unimod1})
\be
\hat{u}_i{}^j (z_{12}) = \left( 
\frac{ x_{ {\bar 2} 1}{}^2 }{x_{ {\bar 1} 2}{}^2} \right)^\hf 
u_i{}^j (z_{12})
\label{unimod2}
\ee 
takes the form
\be
\left(\hat{u}^{-1} (z_{12})\right)_i{}^j
= \hat{u}_i{}^j (z_{21}) 
= \ve^{jk}\, \hat{u}_k{}^l (z_{12})\,\ve_{li}   
\label{unimod3}
\ee
which can be written as
\be 
 \hat{u}_{ji} (z_{21}) ~=~ -\hat{u}_{ij} (z_{12})~.
\label{unimod4}
\ee
The importance of this relation is that it implies that
the two-point function
\bea
A_{i_1 i_2} (z_1, z_2) & \equiv & 
\frac{ \hat{u}_{i_1 i_2} (z_{12})}
{ \left( x_{\bar{1} 2}{}^2   x_{\bar{2} 1}{}^2 \right)^\hf }
= - \frac{ \hat{u}_{i_2 i_1} (z_{21})}
{\left( x_{\bar{1} 2}{}^2   x_{\bar{2} 1}{}^2 \right)^\hf } \non \\
&=& \frac{ u_{i_1 i_2} (z_{12})}
{ x_{\bar{1} 2}{}^2 } =
- \frac{ u_{i_2 i_1} (z_{21})}
{ x_{\bar{2} 1}{}^2 }
\eea
is analytic in $z_1$ and $z_2$ for $z_1 \neq z_2$,
\be
D_{1\,\a (j_1} A_{i_1) i_2} (z_1, z_2) =
{\bar D}_{1\,\ad (j_1} A_{i_1) i_2} (z_1, z_2) = 0~.
\label{import111}
\ee
As we will see later, $A_{i_1 i_2} (z_1, z_2)$ is a building
block of correlation functions of analytic quasi-primary superfields
like the $\cN =2$ flavour current superfields. It is worth noting that 
unitarity of $\hat{u} (z_{12})$ now implies
\be 
\overline{ \hat{u}_{ij} (z_{12}) } 
~=~  \hat{u}^{ij} (z_{12}) ~.
\ee

Above properties of the matrices $\hat{u}_{rs}$, where $r,s =1,2,3$, 
have natural counterparts for ${\bf u} (\bZ_s)$,
with ${\bf u} (\bZ_3)$ defined in (\ref{greatu}).
We introduce the unitary unimodular $2 \times 2$ matrix
\be
\hat{{\bf u}} (\bZ_3)= \left(
 \frac{ {\bar \bX}_3{}^2 }{ \bX_3{}^2 }\right)^\hf \;
{\bf u} (\bZ_3)~, \qquad 
\det \hat{{\bf u}} (\bZ_3)=1~, \qquad 
\hat{{\bf u}}^\dag (\bZ_3) \hat{{\bf u}} (\bZ_3) =1
\ee
with the superconformal transformation law
\be
\d \hat{{\bf u}}_i{}^j (\bZ_3) =
\hat{\L}_i{}^k (z_3)\, \hat{{\bf u}}_k{}^j (\bZ_3) 
- \hat{{\bf u}}_i{}^k (\bZ_3) \,\hat{\L}_k{}^j (z_3)~.
\ee
Since $\hat{{\bf u}} (\bZ_3)$ is unimodular and unitary, 
we 
have
\bea
{\rm tr}\; \hat{{\bf u}}^\dag (\bZ_3) &=&
{\rm tr}\; \hat{{\bf u}} (\bZ_3)\non \\
 \hat{{\bf u}}^\dag_{ji} (\bZ_3) 
&=& - \hat{{\bf u}}_{ij} (\bZ_3)
\eea
and from here one can readily deduce the useful identities
\bea
&&2\left( \frac{1}{ \bar{\bX}_3{}^2 } - 
\frac{1}{ \bX_3{}^2 } \right) =
( \bX_3 \cdot \bar{\bX}_3) 
\left( \frac{1}{ (\bar{\bX}_3{}^2)^2 } - 
\frac{1}{ (\bX_3{}^2)^2 } \right) \non \\
&& \qquad \qquad \Q_{3\,(i} \frac{ \bX_3 }{(\bX_3{}^2)^2}
{\bar \Q}_{3\,j)} =
\Q_{3\,(i} \frac{ \bar{\bX}_3 }{(\bar{\bX}_3{}^2)^2}
{\bar \Q}_{3\,j)} ~.
\label{real4}
\eea

\section{Correlators of \mbox{$\cN$} = 2 quasi-primary superfields}
\setcounter{footnote}{0}

\subsection{Quasi-primary superfields}
In $\cN$-extended superconformal field theory,
a quasi-primary superfield $\cO^\cA_\cI (z)$,
carrying some number of undotted and dotted
spinor indices, denoted collectively by the 
superscript `$\cA$', and transforming in a
representation $T$ of the $R$--symmetry 
SU$(\cN)$ with respect to the subscript `$\cI$'~,
is defined by the following infinitesimal 
transformation law 
under the superconformal group
\bea
\d\, \cO^\cA_\cI (z) &=& - \x \, \cO^\cA_\cI (z) 
+ (\hat{\o}^{\g \d} (z) M_{\g \d}+ 
\hat{ \bar{\o}}^{\dot{\g} \dot{\d}} (z) 
\bar{M}_{\dot{\g} \dot{\d}} )^\cA{}_\cB\,
\cO^\cB_\cI (z) \non \\
&& + {\rm i} \, \hat{\L}^k{}_l (z)\,
(R^l{}_k )_\cI{}^\cJ \cO^\cA_\cJ (z) 
 - 2\left( q\, \s(z) + \bar{q}\, \bar{\s} (z) \right) 
\cO^\cA_\cI (z)~.
\eea
Here $M_{\a \b}$ and $\bar{M}_{\ad \bd}$ are 
the Lorentz generators which act on the undotted 
and dotted spinor indices, respectively, while
$R^i{}_j$ are the generators of SU$(\cN)$.
The parameters $q$ and $\bar q$ determine the dimension  $(q+\bar q)$
and U(1) $R$--symmetry charge $(q-\bar q)$ of the superfield, since 
for a combined scale and U(1) chiral transformation
\be
\d x^m = \l  x^m~, \qquad \d \q^\a_i = 
\hf (\l + {\rm i}\, \O )\q^\a_i ~, \qquad 
\d {\bar \q}^i_\ad =\hf (\l - {\rm i}\, \O )   
{\bar \q}^i_\ad
\ee
we have 
\be
\s (z) = \hf \left( \l + {\rm i}\, 
\frac{\cN}{\cN -4} \O \right)~.
\ee

In this paper we are mainly interested in two- and 
three-point correlation functions of the 
supercurrent $\cJ(z)$ and a flavour current superfield
$\cL_{(ij)} (z) $ in $\cN =2$ superconformal theory.
The reality condition $\bar{ \cJ} =\cJ $ and 
the supercurrent conservation equation
\be
D^{ij} \, \cJ = \bar{D}^{ij} \cJ = 0 
\label{scce2}
\ee
uniquely fix the superconformal transformation 
law of $\cJ$
\be
\d \cJ (z) = -\x\,\cJ (z)   
 - 2\left(  \s(z) +  \bar{\s} (z) \right) \cJ (z)~.
\label{n=2sctrl}
\ee 
As for the flavour current superfield, 
the reality condition $\overline{\cL_{ij} } =
\cL^{ij} $ and the conservation (analyticity) equation
\be
D^{(i}_\a \cL^{jk)} = {\bar D}^{(i}_\ad \cL^{jk)} =0
\label{fcce}
\ee
fix its transformation law to
\be
\d \cL_{ij}(z) = -\x \, \cL_{ij}(z) 
+ 2 {\rm i}\, \hat{\L}_{(i}{}^k (z)\, \cL_{j)k}(z)
- 2\left(  \s(z) +  \bar{\s} (z) \right)\cL_{ij}(z)~.
\label{n=2fcstl}
\ee  
Similar to the $\cN=1$ consideration of \cite{er},
the transformations (\ref{n=2sctrl}) and (\ref{n=2fcstl})
can also be obtained as invariance conditions with 
respect to combined diffeomorphisms and Weyl transformations
in the superconformal theory coupled
to a $\cN =2$ supergravity background.   

\subsection{Two-point functions}  
According to the general prescription of \cite{osborn,park2},
the two-point function of a quasi-primary superfield
$\cO_\cI$ (carrying no Lorentz indices) with its
conjugate $\bar{\cO}^\cJ$ reads
\be
\langle \cO_\cI (z_1)\;\bar{\cO}^\cJ (z_2)\rangle
~=~ C_{\cO}\;\frac{ 
T_\cI{}^\cJ \left( \hat{u}(z_{12}) \right)}
{ (x_{\bar{1}2}{}^2)^{\bar q} (x_{\bar{2}1}{}^2)^q }
\ee
with $ C_{\cO}$ a normalization constant. Here $T$ denotes
the representation of SU$(\cN)$ to which  $\cO_\cI$
belongs.

For the two-point function of the $\cN=2$ supercurrent, 
the above prescription leads to 
\be
\langle \cJ (z_1)\;\cJ (z_2)\rangle 
\;= \; c_\cJ \; \frac{1}{ x_{ {\bar 1} 2}{}^2  
x_{{\bar 2} 1}{}^2}~.
\label{n=2sc,t-pf}
\ee
Using the identity 
\be
\bar{D}_1{}^{ij}\, \frac{1}{x_{\bar{1}2}{}^2}~=~  
4{\rm i}\, D_1{}^{ij} \, \d^8_+ (z_1, z_2)~,
\ee
where $\d^8_+ (z_1, z_2)$ denotes the $\cN=2$
chiral delta function,
\be
\d^8_+ (z_1, z_2) ~=~ \frac{1}{16}\bar{D}^4 \,
\d^{12}(z_1 -z_2)~,\qquad 
\bar{D}^4~=~ \frac{1}{3}\bar{D}^{ij}\bar{D}_{ij}~,
\ee
we immediately see that the supercurrent
conservation equation is satisfied at non-coincident points
\be
D_1{}^{ij} \langle \cJ (z_1)\;\cJ (z_2) \rangle~=~
\bar{D}_1{}^{ij}  \langle \cJ (z_1)\;\cJ (z_2) \rangle
\; =\; 0 ~, \qquad z_1 \neq z_2 ~.
\ee
In this paper we leave aside the analysis
of singular behaviour at coincident points, see
\cite{osborn} for details.

In the case of 
two-point function of the $\cN=2$  flavour current 
superfield $\cL_{ij}$, the above prescription gives
\be
\langle \cL_{i_1 j_1} (z_1) \cL^{i_2 j_2} (z_2)\rangle \;=\;
c_\cL \;
\frac{
\hat{u}_{i_1}{}^{i_2}(z_{12})  \hat{u}_{j_1}{}^{j_2}(z_{12}) + 
\hat{u}_{i_1}{}^{ j_2}(z_{12})  \hat{u}_{j_1}{}^{ i_2} (z_{12}) }
{  x_{ {\bar 1} 2}{}^2  x_{{\bar 2} 1}{}^2} ~.
\label{n=2fcst-pf}
\ee
Because of eq. (\ref{import111}), the relevant
conservation equation is satisfied
\be
D_{1\, \a (k_1} \langle \cL_{i_1 j_1)} (z_1) \cL^{i_2 j_2} (z_2)\rangle 
\;=\; \bar{D}_{1\, \ad (k_1} 
\langle \cL_{i_1 j_1)} (z_1) \cL^{i_2 j_2} (z_2)\rangle 
~=~ 0 
\ee
for $z_1 \neq z_2$.
 
\subsection{Three-point functions}
According to the general prescription of 
\cite{osborn,park2}, the three-point function 
of quasi-primary superfields $\cO^{(1)}_{\cI_1}$,
$\cO^{(2)}_{\cI_2}$ and $\cO^{(3)}_{\cI_3}$
reads
\be
\langle
\cO^{(1)}_{\cI_1} (z_1)\, \cO^{(2)}_{\cI_2} (z_2)\,
\cO^{(3)}_{\cI_3} (z_3)
\rangle = 
\frac{ 
T^{(1)}{}_{\cI_1}{}^{\cJ_1} \left( \hat{u}(z_{13}) \right)
T^{(2)}{}_{\cI_2}{}^{\cJ_2} \left( \hat{u}(z_{23}) \right)
}
{ 
(x_{\bar{1}3}{}^2)^{\bar{q}_1} (x_{\bar{3}1}{}^2)^{q_1} 
(x_{\bar{2}3}{}^2)^{\bar{q}_2} (x_{\bar{3}2}{}^2)^{q_2}
}
H_{\cJ_1 \cJ_2 \cI_3} (\bZ_3)~.
\ee 
Here $H_{\cJ_1 \cJ_2 \cI_3} (\bZ_3)$ transforms
as an isotensor at $z_3$ in the representations
$T^{(1)},~T^{(2)}$ and $T^{(3)}$ with respect 
to the indices $\cJ_1,~ \cJ_2$ and $\cI_3$, respectively,
and possesses the homogeneity property
\bea
&& H_{\cJ_1 \cJ_2 \cI_3} ( \D \bar{\D}\,\bX,
\D\, \Q, \bar{\D} {\bar \Q}) =
\D^{2a} \bar{\D}^{2\bar{a}}
H_{\cJ_1 \cJ_2 \cI_3} ( \bX, \Q, {\bar \Q}) \non \\
&& 
\Big( \frac{2}{\cN} - 1\Big) a - \frac{2}{\cN} \bar a =\bar{q}_1 + \bar{q}_2 -q_3~,\qquad
\Big( \frac{2}{\cN} - 1\Big) \bar a - \frac{2}{\cN}  a ={q}_1 + {q}_2 - \bar{q}_3~,
\label{3.72}
\eea
which we use only  for the case where $q_i=\bar{q}_i$.\footnote{We thank Hongliang Jiang for pointing out an error in  eq. (\ref{3.72}) as given in the previous versions of the manuscript, as well as Benjamin Stone for re-deriving the correct homogeneity relations. However,  for the case of our interest in this paper, $q_i=\bar{q}_i$, the subsequent results are unmodified.} 
In general, the latter equation admits
a finite number of linearly independent solutions,
and this can be considerably reduced by taking into
account the symmetry properties, superfield conservation
equations and, of course, the superfield constraints
(chirality or analyticity).

\subsubsection{The \mbox{$\cN$} = 2 supercurrent}

We are going to analyse the three-point function 
of the $\cN=2$ supercurrent for which we should have
\bea
\langle
\cJ (z_1)\, \cJ (z_2)\, \cJ (z_3)
\rangle & = &
\frac{1}{ 
x_{\bar{1}3}{}^2 x_{\bar{3}1}{}^2 
x_{\bar{2}3}{}^2 x_{\bar{3}2}{}^2}
H (\bZ_3)~,   
\label{sc3-pf0} \\ 
H ( \D \bar{\D}\,\bX,
\D\, \Q, \bar{\D} {\bar \Q}) &=&
(\D \bar{\D})^{-2}
H ( \bX, \Q, {\bar \Q}) \non
\eea
where the real function $H (\bZ_3)$ has to be compatible with the 
supercurrent conservation equation and the 
symmetry properties with respect to transposition
of indices. Since $H (\bZ_3)$ is invariant under
U$(1)\times {\rm SU}(2)$ $R$--transformations,
we have $H(\bX_3, \Q_3, \bar{\Q}_3) = H' (\bX_3 , \bar{\bX}_3)$,
as a consequence of (\ref{bosferm}). 

When analysing the restrictions imposed by the 
$\cN = 2$ conservation equations,
it proves advantageous, following similar $\cN  = 1$ 
considerations in \cite{osborn}, 
to make use of conformally covariant operators
$\cD_{\bar A} = (\pa / \pa \bX^a_3, \cD_{\a i}, \bar{\cD}^{\ad i}) $
and $\cQ_{\bar A} = (\pa / \pa \bX^a_3, {\cal Q}_{\a i},
\bar{\cal Q}^{\ad i})$ defined by 
\bea
  \cD_{\a i} &=& \frac{\pa}{ \pa \Q^{\a i}_3 }
-2{\rm i}\,(\s^a)_{\a \ad} \bar{\Q}^\ad_{3\, i}
\frac{\pa }{ \pa \bX^a_3 }~, \qquad
\bar{\cD}^{\ad i} = \frac{\pa}{ \pa \bar{\Q}_{3\, \ad i} } 
\non \\
{\cal Q}_{\a i}  &=&  \frac{\pa}{ \pa \Q^{\a i}_3 }~, \qquad
\bar{\cal Q}^{\ad i} \frac{\pa}{ \pa \bar{\Q}_{3\, \ad i} } 
+ 2{\rm i}\, \Q^i_{3\, \a} (\tilde{\s}^a)^{\ad \a}
\frac{\pa}{ \pa \bX^a_3 }  \non \\
&& \qquad \qquad  [\cD_{\bar A}  ,  \cQ_{\bar B} \} ~=~0 ~. 
\label{d-q}
\eea
These operators emerge via the relations
\bea
D_{1\,\a}{}^i \;t (\bX_3, \Q_3, \bar{\Q}_3) &=&
-{\rm i} \,(\tilde{x}_{\bar{3} 1}{}^{-1})_{\a \bd}\;
u_j{}^i(z_{31})\; \bar{\cD}^{\bd j}\;
 t (\bX_3, \Q_3, \bar{\Q}_3) \non \\
\bar{D}_{1\,\a i} \; t (\bX_3, \Q_3, \bar{\Q}_3) &=&
-{\rm i}\, (\tilde{x}_{\bar{1} 3}{}^{-1})_{\b \ad}\;
u_i{}^j(z_{13})\; \cD^b_j \;t (\bX_3, \Q_3, \bar{\Q}_3)\non \\
D_{2\,\a}{}^i \;t (\bX_3, \Q_3, \bar{\Q}_3) &=&
~~{\rm i} \,(\tilde{x}_{\bar{3} 2}{}^{-1})_{\a \bd}\;
u_j{}^i(z_{32})\; \bar{\cQ}^{\bd j}\;
 t (\bX_3, \Q_3, \bar{\Q}_3) \non \\
\bar{D}_{2\,\a i}\; t (\bX_3, \Q_3, \bar{\Q}_3) &=&
~~{\rm i}\, (\tilde{x}_{\bar{2} 3}{}^{-1})_{\b \ad}\;
u_i{}^j(z_{23}) \;\cQ^b_j \;t (\bX_3, \Q_3, \bar{\Q}_3)
\label{d-q2}
\eea
where $t (\bX_3, \Q_3, \bar{\Q}_3)$ is an arbitrary function.

With the aid of these operators, 
one can prove the identity
\be 
D_1{}^{\a i} D_{1\,\a}{}^j 
\left( \frac{1}{ x_{\bar{3} 1}{}^2 }
t (\bX_3, \Q_3, \bar{\Q}_3) \right) ~=~
- \frac{ u^i{}_k (z_{13}) \, u^j{}_l (z_{13})}
{(x_{\bar{1}3}{}^2)^2}  \;
\bar{\cD}_\ad^k \, \bar{\cD}^{\ad l}\;
t (\bX_3, \Q_3, \bar{\Q}_3)
\ee
and similar ones involving the operators
$\bar{D}_{1\, ij}$, $D_2{}^{ij}$ and 
$\bar{D}_{2\, ij}$.

Now, the supercurrent conservation equation (\ref{scce2})
leads to the requirements
\be
\cD_{ij}\,H (\bX_3, \Q_3, \bar{\Q}_3) 
={\bar \cD}^{ij}\,H (\bX_3, \Q_3, \bar{\Q}_3) =0
\label{scceq3}
\ee
and to similar ones with $\cD$'s $\longrightarrow$ $\cQ$'s.
Since $\bar{\cD}^{\ad i}$ and $\cQ_{\a i}$ coincide with partial 
fermionic derivatives the above equations imply 
\be
\frac{\pa^2}{ \pa \Q_{\a}^{ i} \, \pa \Q^{\a j} } 
H (\bX, \Q, \bar{\Q}) = 
\frac{\pa^2}{\pa \bar{\Q}^\ad_i \, \pa \bar{\Q}_{\ad j} } 
H (\bX, \Q, \bar{\Q})
=0~,
\ee 
and therefore the power series
of $H (\bX, \Q, \bar{\Q})$
in the Grassmann variables $\Q$'s 
contains only few terms.

The general solution for $H(\bZ_3)$ compatible
with all the physical requirements on the three-point
function of the $\cN=2$ supercurrent reads
\be
H (\bX_3, \Q_3, \bar{\Q}_3) =
A \, \left(\frac{1}{ \bX_3{}^2} + \frac{1}{{\bar \bX}_3{}^2 }\right) 
+B\, \frac{ 
\Q_3^{\a \b} \bX_{3 \a \ad} \bX_{3 \b \bd} 
{\bar \Q}_3^{\ad \bd} }
{(\bX_3{}^2)^2}  
\label{sc3-pf}
\ee
where
\be
\Q_3^{\a \b} ~=~ \Q_3^{(\a \b)}~=~
\Q_3^{\a i} \, \Q_{3\,i}^\b~, \qquad \quad
\bar{\Q}_3^{\ad \bd} ~=~ \bar{\Q}_3^{(\ad \bd)}~=~
\bar{\Q}_{3\,i}^\ad \, \bar{\Q}_3^{\ad i}
\ee
and  $A,~B $ are real parameters. Note that the second 
structure is nilpotent. 

Let us comment on the derivation of this solution.
First, it is straightforward to check that
the functions $\bX_3{}^{-2}$ and $\bar{\bX}_3{}^{-2}$
satisfy eq. (\ref{scceq3}). They enter $H(\bZ_3)$
with the same real coefficient, since $H$ must be real
and invariant under the replacement 
$z_1 ~\leftrightarrow~z_2$ that acts on $\bX_3$ and
$\bar{\bX}_3$ by $\bX_3~\leftrightarrow~ - \bar{\bX}_3$.
The second term in (\ref{sc3-pf}) is a solution to 
(\ref{scceq3}) due to the special $\cN =2$ identity
\be
D^{ij} \; \q^{\ab} ~=~ \bar{D}^{ij}\; \bar{\q}^{\ad \bd} ~=~0~.
\ee

It is important to demonstrate that the second 
term  in (\ref{sc3-pf}) is real, i.e. that  
\be
\frac{ 
\Q_3^{\a \b} \bX_{3 \a \ad} \bX_{3 \b \bd} 
{\bar \Q}_3^{\ad \bd} }
{(\bX_3{}^2)^2} ~=~
\frac{ 
\Q_3^{\a \b} \bar{\bX}_{3 \a \ad} \bar{\bX}_{3 \b \bd} 
{\bar \Q}_3^{\ad \bd} }
{(\bar{\bX}_3{}^2)^2}~.
\label{realbstr}
\ee 
Using the identity $\bar{\bX}_3^a = \bX_3^a 
+ 2 {\rm i}\,
\Q_3^i\, \s^a \,{\bar \Q}_{3\,i}$, we first represent
$\bar{\bX}_3{}^{-2}$ as a function of $\bX_3,\Q_3, \bar{\Q}_3$:
\be
\frac{1}{\bar{\bX}_3{}^2} = \frac{1}{\bX_3{}^2}
- 4{\rm i}\, \frac{ \Q_3^i \bX_3 \bar{\Q}_{3 i} }
{(\bX_3{}^2)^2}
-8 \frac{ 
\Q_3^{\a \b} \bX_{3\, \a \ad}  \bX_{3\, \b \bd} 
\bar{\Q}_3^{\ad \bd} }
{(\bX_3{}^2)^3}~.
\label{twox}
\ee 
We then apply the same identity to express 
$\Q_3^i \bX_3 \bar{\Q}_{3 i}$ in
the second term  via $\bX_3$ and ${\bar \bX}_3$.
Now, eq. (\ref{realbstr}) follows from (\ref{twox})
and the first identity in (\ref{real4}).

Using (\ref{difz}), one can check that the three-point 
function (\ref{sc3-pf0}), (\ref{sc3-pf}) is completely symmetric
in its arguments. 
 
\subsubsection{Flavour current superfields}
Let us turn to the three-point function of flavour current 
superfields $\cL^{\abar}_{ij}$  
\bea
&&\langle
\cL^{\abar}_{i_1 j_1} (z_1)\,  \cL^{\bbar}_{i_2 j_2} (z_2)\,
\cL^{\cbar}_{i_3 j_3} (z_3)
\rangle  \label{abc} \\
&&\qquad\qquad\qquad =\frac{ 
\hat{u}_{i_1}{}^{k_1} (z_{13})  \hat{u}_{j_1}{}^{l_1} (z_{13}) 
\hat{u}_{i_2}{}^{k_2} (z_{23})  \hat{u}_{j_2}{}^{l_2} (z_{23})
}{
x_{ {\bar 3} 1}{}^2  x_{{\bar 1} 3}{}^2
x_{ {\bar 3} 2}{}^2  x_{{\bar 2} 3}{}^2 } 
\cdot
H^{{\abar} {\bbar} {\cbar} }_{(k_1 l_1)(k_2 l_2) (i_3 j_3)} 
(\bZ_3)~, \non
\eea
with
\be
H^{{\abar} {\bbar} {\cbar} }_{(k_1 l_1)(k_2 l_2)(i_3 j_3)} 
( \D \bar{\D}\,\bX,
\D\, \Q, \bar{\D} {\bar \Q}) =
(\D \bar{\D})^{-2} 
H^{{\abar} {\bbar} {\cbar} }_{(k_1 l_1)(k_2 l_2)(i_3 j_3)} 
(\bX, \Q, {\bar \Q})~.
\ee
Using relations (\ref{import111}) and (\ref{d-q2}),
the flavour current conservation equations (\ref{fcce})
are equivalent to 
\bea 
\cD_{\a\,(i_1} H^{{\abar} {\bbar} {\cbar} }_{k_1 l_1)(k_2 l_2)(i_3 j_3)} 
(\bX, \Q, {\bar \Q}) &=& 
\bar{\cD}_{\ad\, (i_1}
H^{{\abar} {\bbar} {\cbar} }_{k_1 l_1)(k_2 l_2)(i_3 j_3)}
(\bX, \Q, {\bar \Q}) = 0~,\non \\
\cQ_{\a\,(i_2} 
H^{{\abar} {\bbar} {\cbar} }_{|k_1 l_1|k_2 l_2)(i_3 j_3)} 
(\bX, \Q, {\bar \Q}) &=& 
\bar{\cQ}_{\ad\, (i_2}
H^{{\abar} {\bbar} {\cbar} }_{|k_1 l_1|k_2 l_2)(i_3 j_3)} 
(\bX, \Q, {\bar \Q}) = 0~.
\eea
In particular, since 
$\bar{\cD}^{\ad i}$ and $\cQ_{\a i}$ are just partial
Grassmann derivatives, we should have
\be
\frac{\pa}{\pa \bar{\Q}_{3\,\ad}{}^{(i_1} }
H^{{\abar} {\bbar} {\cbar} }_{k_1 l_1)(k_2 l_2)(i_3 j_3)}
(\bX, \Q, {\bar \Q}) =
\frac{\pa}{\pa \Q_{3 \, \a}{}^{(i_2} }
H^{{\abar} {\bbar} {\cbar} }_{|k_1 l_1|k_2 l_2)(i_3 j_3)} 
(\bX, \Q, {\bar \Q}) = 0~.
\ee 
The most general form for the correlation function in question
is of the form (\ref{abc}) with
\be
H^{{\abar} {\bbar} {\cbar} }_{(k_1 l_1)(k_2 l_2) (i_3 j_3)} 
(\bZ_3) = f^{{\abar} {\bbar} {\cbar} } \;
\frac{ \ve_{ i_3(k_1 } \hat{{\bf u}}{}_{\,l_1)(l_2} (\bZ_3 )
\ve_{ k_2)j_3 } } { (\bX_3{}^2 \bar{\bX}_3{}^2 )^\hf} \quad 
+ \quad (i_3 \longleftrightarrow j_3)
\label{n=2fcs3p-f2}
\ee
with $f^{\abar \bbar \cbar}=
f^{[\abar \bbar \cbar]}$ a completely antisymmetric tensor, 
proportional to the structure constants of the flavour group.
In contrast to $\cN=1$ supersymmetry \cite{osborn}, the three-point 
correlation function of flavour currents
does not admit an anomalous term proportional to an overall completely
symmetric group tensor, 
$d^{\abar \bbar \cbar}= d^{(\abar \bbar \cbar)}$.
This is a consequence of the fact that the $\cN =2$
conservation equations (\ref{fcce})
do not admit non-trivial deformations; see also below.

\subsubsection{Mixed correlators}
The three-point function involving two $\cN=2$ supercurrent
insertions and a flavour $\cN=2$ current superfield, 
turns out to vanish
\be
\langle
\cJ (z_1) \; \cJ (z_2) \; \cL_{ij} (z_3)
\rangle ~=~ 0\;.
\ee 
On general grounds, the only possible expression
for such a correlation function compatible with 
the conservation equations and reality 
properties should read
\be
\langle
\cJ (z_1) \; \cJ (z_2) \; \cL_{ij} (z_3)
\rangle ~=~
P \frac{\rm 1}{
x_{ {\bar 3} 1}{}^2  x_{{\bar 1} 3}{}^2
x_{ {\bar 3} 2}{}^2  x_{{\bar 2} 3}{}^2 }\;
\frac{ \hat{{\bf u}}_{(ij)} (\bZ_3 )}
{( \bX_3{}^2 \bar{\bX}_3{}^2 )^\hf }
\ee
with $P$ a real constant. However, the hight-hand-side
is easily seen to be antisymmetric with respect 
to the transposition $z_1 \leftrightarrow z_2$
acting as $\bX_3  \leftrightarrow  - \bar{\bX}_3$,
and hence
$\hat{{\bf u}}_{(ij)} (\bZ_3 ) ~ \leftrightarrow ~
\hat{{\bf u}}^\dag_{(ij)} (\bZ_3 ) = - 
\hat{{\bf u}}_{(ij)} (\bZ_3 )$.
Therefore, we must set $P=0$.  

{}For the three-point function with two flavour currents and 
one supercurrent insertion one finds
\bea
&&\langle
\cL^{\abar}_{i_1 j_1} (z_1)\,  \cL^{\bbar}_{i_2 j_2} (z_2)\,
\cJ (z_3)
\rangle  \label{l-l-j} \\
&& \qquad =  d\;\d^{{\abar} {\bbar} } \; \frac{ 
\hat{u}_{i_1}{}^{k_1} (z_{13})  \hat{u}_{j_1}{}^{l_1} (z_{13}) 
\hat{u}_{i_2}{}^{k_2} (z_{23})  \hat{u}_{j_2}{}^{l_2} (z_{23})
}{
x_{ {\bar 3} 1}{}^2  x_{{\bar 1} 3}{}^2
x_{ {\bar 3} 2}{}^2  x_{{\bar 2} 3}{}^2 } 
\cdot
\frac{ \ve_{ k_2(k_1 } \hat{{\bf u}}{}_{\,l_1)l_2} (\bZ_3 )
+ \ve_{ l_2(k_1 } \hat{{\bf u}}{}_{\,l_1)k_2} (\bZ_3 )
} { (\bX_3{}^2 \bar{\bX}_3{}^2 )^\hf} ~.\non
%\label{l-l-j}
\eea
with $d$ being a real parameter which can be 
related, via supersymmetric Ward identities, 
to the parameter $c_\cL$ in the two-point
function (\ref{n=2fcst-pf}).  

\subsection{Example: \mbox{$\cN$} = 4 super Yang-Mills}
Let us consider the harmonic superspace formulation for 
$\cN=4$ super Yang-Mills theory
\be
S[ V^{++}, q^+, \breve{q}^+ ] ~= ~
\frac{1}{2g^2} \,{\rm tr} \int {\rm d}^4 x\,
{\rm d}^4 \q  \,  W^2
-  \frac{1}{g^2} \,{\rm tr}  \int {\rm d}u\, {\rm d} \z^{(-4)}\, 
\breve{q}^+ \left( D{}^{++}
+ {\rm i}\, V^{++}  \right) q^+\;.
\ee
Since the hypermultiplet $q^+$ belongs to the adjoint representation 
of the gauge group, we can unify $q^+$ and $\breve{q}^+$
in an isospinor
\be
q^+_{\overline{\imath}} 
= (q^+, \breve{q}^+)~, \qquad
q^{+\,\overline{\imath}}= 
\ve^{\overline{\imath} \overline{\jmath}}\,
q^+_{\overline{\jmath}} = (\breve{q}^+ , -q^+)~,
\qquad (q^+_{\overline{\imath}})\, \breve{} 
= q^{+\,\overline{\imath}}
\ee
such that the action takes the form (with 
$\nabla =  D{}^{++}+ {\rm i} \, V^{++}$)
\be
S ~= ~
\frac{1}{2g^2} \,{\rm tr} \int {\rm d}^4 x\,
{\rm d}^4 \q  \,  W^2
-  \frac{1}{2g^2} \,{\rm tr}  \int {\rm d}u\, {\rm d} \z^{(-4)}\, 
\breve{q}^{+\,\overline{\imath}} 
\stackrel{\longleftrightarrow}{\nabla}{}^{--}
q^+_{\overline{\imath}}
\ee
This form makes it explicit that the theory 
manifestly possesses
the flavour symmetry SU${}_F(2)$, in addition to the 
$\cN=2$ automorphism group SU${}_R(2) \times {\rm U}_R(1)$.
The full group SU${}_R(2) \times {\rm U}_R(1) 
\times {\rm SU}_F(2)$ is the maximal subgroup of ${\rm SU}_R(4)$ --
the $R$--symmetry group of the $\cN =4$ SYM -- which can be made manifest
in the framework of $\cN = 2$ superspace formulation.
While the conserved currents for SU${}_R(2) \times {\rm U}_R(1)$ 
belong to the supercurrent 
\be
\cJ ~=~ \frac{1}{g^2} \,{\rm tr} \left( 
{\bar W} W 
~-~\frac{1}{4}\,
q^{+\, \overline{\imath} }   
\stackrel{\longleftrightarrow}{\nabla}{}^{--}
q^+_{ \overline{\imath} } \right)~,
\ee
the currents for SU${}_F(2)$ belong to the flavour
current supermultiplet 
\be
\cL^{++\,\overline{a}}(z,u)~\propto ~
{\rm i}\, q^{+\, \overline{\imath} } \,
(\t^{\overline{a}})_{\overline{\imath}}\,{}^{\overline{\jmath}}\,
q^+_{ \overline{\jmath} }  
~=~ u^{+i} u^{+i}\,
\cL^{\overline{a}}_{ij} (z)~, 
\label{n=4fc}
\ee
with $\t^{\overline{a}}$ the Pauli matrices; 
here the latter equality is valid on shell.
The fact that $\langle \cJ\cJ\cL\rangle$ vanishes identically whereas 
$\langle \cL\cL\cJ\rangle$ is generically non-zero is now a simple consequence of 
group theory. In fact, group theory restricts the
structure of the correlation function 
of three $\cN=4$ SU$_R(4)$ currents to be proportional to 
${\rm tr}\,(t^I t^J t^K)$ where $t^I$ is a SU$_R(4)$ generator. 
By considering the action of the $\cN=2$ U$_R(1)$ symmetry, one finds that the 
correct embedding u$_R(1)\subset {\rm su}_R(4)$ is ${\rm diag}(+1,+1,-1,-1)$.  
Also 
${\rm su}_R(2) \bigoplus {\rm su}_F(2)\subset{\rm su}_R(4)$ is embedded as
${\rm diag}({\rm su}_R(2),{\rm su}_F(2))$. The result stated above now follows 
immediately. Three- and four- point functions 
of the flavour currents (\ref{n=4fc}) have been computed at two loops
in \cite{HSSW}.

\section{Reduction to \mbox{$\cN$} = 1 superfields}
\setcounter{footnote}{0}
{}From the point of view of 
$\cN=1$ superconformal symmetry,
any $\cN=2$ quasi-primary superfield
consists of several $\cN=1$ quasi-primary superfields. 
Having computed the correlation functions
of $\cN=2$ quasi-primary superfields,
one can read off all correlators
of their $\cN=1$ superconformal components.

When restricting ourselves to the subgroup ${\rm SU}(2,2|1) \in
{\rm SU}(2,2|2)$, all matrix elements of ${\bf h}(z)$
(\ref{stability}) with $i,j= \underline{2}$ should vanish,
and hence we have to set
\be
\hat{\L}_{\underline{1}}{}^{\underline{2}}=
\hat{\L}_{\underline{2}}{}^{\underline{1}}= 0~, \qquad
{\rm i}\, \hat{\L}_{\underline{1}}{}^{\underline{1}}= 
- {\rm i}\,\hat{\L}_{\underline{2}}{}^{\underline{2}}=
\bar{\s} - \s ~.
\label{n=1R-sym}
\ee
Therefore, the $\cN =1$ U(1) $R$-transformation 
is a combination of $\cN =2$ U(1)   
and special SU(2) $R$-transformations. 

Keeping eq. (\ref{n=1R-sym}) 
in mind, from the $\cN=2$ supercurrent transformation 
law (\ref{n=2sctrl}) one deduces the transformation
of the $\cN =1$ currents (\ref{n=1s-ccomponents}) 
\bea
\d J  &=& -\x\,J    
 - 2\left(  \s  +  \bar{\s}  \right) J  \non \\
\d J_\a  &=& - \x\,J_\a  +\hat{\o}_\a{}^\b 
 J_\b  - \left(3 
\s +  2 \bar{\s}  \right) J_\a  \non \\ 
\d J_{\a \ad}  &=& - \x\,J_{\a \ad} 
+ ( \hat{\o}_\a{}^\b  \d_\ad{}^\bd 
+ \bar{\hat{\o}}_\ad{}^\bd  \d_\a{}^\b) J_{\b \bd} 
- 3\left(  
\s +  \bar{\s}  \right) J_{\a \ad} ~.
\eea
These superconformal transformations 
are uniquely singled out by the relevant
conservation equations
\bea
D^2\, J &=& \bar{D}^2 \,J ~=~ 0 \non \\
D^\a \, J_\a & = &\bar{D}^2 \,J_\a ~=~ 0\non \\
D^\a \, J_{\a  \ad} & = & \bar{D}^\ad \,J_{\a  \ad}
~=~0~.
\eea
In the case of $\cN=2$ flavour current superfield $\cL^{ij}$,
its most interesting $\cN=1$ component 
containing the conserved current, 
\be
L ~ \equiv ~{\rm i}\, \cL^{ \underline{12} }| ~=~\bar{L}
\label{n=1fcs}
\ee
satisfies the standard $\cN=1$ conservation equation
\be
D^2\, L ~=~\bar{D}^2\, L ~=~0~.
\ee
and, therefore, its superconformal transformation 
rule is similar to $J$,
\be
\d L  = -\x\,L    
 - 2\left(  \s  +  \bar{\s}  \right) L~. 
\ee
The same $\cN=1$ transformation follows from
(\ref{n=2fcstl}).

\subsection{Two-point functions}
Using the explicit form (\ref{n=2sc,t-pf}) 
of the $\cN=2$ supercurrent two-point function,
one can read off the two-point functions of the 
$\cN=1$ quasi-primary superfields contained in 
$\cJ$\footnote{Here and below, all building blocks are 
expressed in $\cN=1$ superspace.}
\bea
\langle
J (z_1)\;J (z_2)
\rangle 
&=&  c_\cJ \; \frac{1}{ x_{ {\bar 1} 2}{}^2  
x_{{\bar 2} 1}{}^2}\,, \non \\
\langle
J_\a (z_1)\;\bar{J}_\bd (z_2)
\rangle
 &=& 4{\rm i}\,c_\cJ \;
\frac{ (x_{1\bar{2}})_{\a \bd}}{ x_{ {\bar 1} 2}{}^2  
(x_{{\bar 2} 1}{}^2)^2}\,,  \non \\
\langle
J_{\a \ad}(z_1)\;J_{\b \bd} (z_2)
\rangle
&=&  \frac{64}{3}\, c_\cJ \;
\frac{ (x_{1\bar{2}})_{\a \bd}
(x_{2\bar{1}})_{\b \ad}}{( x_{ {\bar 1} 2}{}^2  
x_{{\bar 2} 1}{}^2)^2}\,. 
\label{n=1sct-pf}
\eea
These results are in agreement with 
$\cN=1$ superconformal considerations \cite{osborn}. 
Similarly, the two-point function of the $\cN=1$
flavour current superfield (\ref{n=1fcs}) follows
from (\ref{n=2fcst-pf})
\be
\langle
L (z_1) \, L (z_2)
\rangle  \;=\;
c_\cL \;
\frac{1 }
{  x_{ {\bar 1} 2}{}^2  x_{{\bar 2} 1}{}^2} ~.
\ee

\subsection{Three-point functions}
We now present several $\cN=1$ three-point functions
which are encoded in that of the $\cN=2$ supercurrent,
given by eqs.
(\ref{sc3-pf0}) and (\ref{sc3-pf}). First of all,
for the leading $\cN =1$ component of $\cJ$
one immediately gets
\be
\langle
J (z_1)\, J (z_2)\, J (z_3)
\rangle  ~= ~
\frac{A}{ 
x_{\bar{1}3}{}^2 x_{\bar{3}1}{}^2 
x_{\bar{2}3}{}^2 x_{\bar{3}2}{}^2}
\left(\frac{1}{ \bX_3{}^2} + \frac{1}{{\bar \bX}_3{}^2 }\right) 
\ee
The second term in (\ref{sc3-pf}) does not contribute
to this three-point function, since $\Q^{\a \b}$ is equal to zero
for $\q_{\underline{2}} = \bar{\q}^{\underline{2}} =0$.

The derivation of three-point functions involving the $\cN=1$ 
supercurrent is technically more complicated.
In accordance with eq. (\ref{n=1s-ccomponents}), 
$J_{\a \ad} $ is obtained from $\cJ$ by applying the
operator
\be
\D_{\a \ad}~=~ \hf [D^{ \underline{2} }_\a \, , \, 
{\bar D}_{\ad \underline{2} }] 
-\frac{1}{6}  [D^{ \underline{1} }_\a \, , \, 
{\bar D}_{\ad \underline{1} }] 
\label{scoperator}
\ee
and, then, the Grassmann variables
$\q_{\underline{2}}$ and $\bar{\q}^{\underline{2}}$
have to be switched off. One can 
prove the following useful relations
\bea
\langle
J_{\a \ad} (z_1)\, J (z_2)\, J (z_3)
\rangle & = &
\frac{
(\tilde{x}_{1 \bar{3}}{}^{-1})_{\a \dot{\g}}
(\tilde{x}_{3 \bar{1}}{}^{-1})_{\g \dot{\a}}
}{ 
x_{\bar{1}3}{}^2 x_{\bar{3}1}{}^2 
x_{\bar{2}3}{}^2 x_{\bar{3}2}{}^2}\;
\D_{(\cD)}{}^{\g \dot{\g}}\; H (\bZ_3)|~, \non \\ 
\langle
J_{\a \ad} (z_1)\, J_{\b \bd} (z_2)\, J (z_3)
\rangle & = &
\frac{
(\tilde{x}_{1 \bar{3}}{}^{-1})_{\a \dot{\g}}
(\tilde{x}_{3 \bar{1}}{}^{-1})_{\g \dot{\a}}
(\tilde{x}_{2 \bar{3}}{}^{-1})_{\b \dot{\d}}
(\tilde{x}_{3 \bar{2}}{}^{-1})_{\d \dot{\b}}
}{ 
x_{\bar{1}3}{}^2 x_{\bar{3}1}{}^2 
x_{\bar{2}3}{}^2 x_{\bar{3}2}{}^2} \non \\ 
& \times & \D_{(\cD)}{}^{\g \dot{\g}}
\: \D_{(\cQ)}{}^{\d \dot{\d}}
\; H (\bZ_3)|
\eea 
where $H (\bZ_3)$ is given by eq. (\ref{sc3-pf}) and
the operators $\D_{(\cD)}$ and $\D_{(\cQ)}$
are constructed in terms of the conformally covariant 
derivatives (\ref{d-q})
\be
\D_{(\cD)}{}^{ \a \ad } = \hf [\cD_{ \underline{2} }^\a \, , \, 
{\bar \cD}^{\ad \underline{2} }] 
-\frac{1}{6}  [\cD_{ \underline{1} }^\a \, , \, 
{\bar \cD}^{\ad \underline{1} }]~, \qquad
\D_{(\cQ)}{}^{ \a \ad } = \hf [\cQ_{ \underline{2} }^\a \, , \, 
{\bar \cQ}^{\ad \underline{2} }] 
-\frac{1}{6}  [\cQ_{ \underline{1} }^\a \, , \, 
{\bar \cQ}^{\ad \underline{1} }] ~.
\ee
Direct calculations lead to 
\bea
\langle
J (z_1)\, J (z_2) \,J_{\a \ad} (z_3) 
\rangle & = &
- \frac{1 }{12} (8 A - 3B)\;
\frac{1}{ 
x_{\bar{1}3}{}^2 x_{\bar{3}1}{}^2 
x_{\bar{2}3}{}^2 x_{\bar{3}2}{}^2 } \non \\
& \times & \left( 
\frac{ 2 (\bP_3 \cdot \bX_3) \bX_{3\, \a \ad} 
+ \bX_3{}^2 \bP_{3\, \a \ad}  }
{(\bX_3{}^2)^2 } 
~+~(\bX_3  \leftrightarrow  - \bar{\bX}_3)
%\frac{ \bX_{3\, \a \ad} }{ \bar{\bX}_3{}^2 }
%- \frac{ \bar{\bX}_{3\, \a \ad} }{\bX_3{}^2}
\right)~, 
\label{JJ-sc} \\
%\noalign{\vskip.3cm}
\langle
J_{\a \ad} (z_1)\, J_{\b \bd} (z_2)\, J (z_3)
\rangle & = &
-\frac{4}{9}(8 A +3 B)\;
\frac{
(x_{1 \bar{3}})_{\a \dot{\g}}
(x_{3 \bar{1}})_{\g \dot{\a}}
(x_{2 \bar{3}})_{\b \dot{\d}}
(x_{3 \bar{2}})_{\d \dot{\b}}
}{ 
(x_{\bar{1}3}{}^2 x_{\bar{3}1}{}^2 
x_{\bar{2}3}{}^2 x_{\bar{3}2}{}^2)^2 } \non \\
& \times & 
\left(
\frac{ 
\bX_3{}^{ \g \dot{\g} }  \bX_3{}^{ \d \dot{\d} }   }
{ (\bX_3{}^2)^3 } +\hf \,
\frac{ 
{\ve}^{\g \d}\; {\ve}^{ \dot{\g} \dot{\d} } }
{(\bX_3{}^2)^2 } 
~+~(\bX_3  \leftrightarrow  - \bar{\bX}_3) 
\right)~,
\label{sc-sc-fcs}
\eea
with $\bP_a$ defined by  \cite{osborn}
\be
{\bar \bX}_a - \bX_a = {\rm i}\, \bP_a ~, \qquad 
\bP_a = 2\, \Q \s_a \bar \Q ~.
%\bP_a \bP_b = \frac{1}{4}\, \eta_{ab}\,  \bP^2~,\qquad
%\bP^2 = - 8\, \Q^2 \, \bar \Q^2~.
\ee

Eq. (\ref{sc-sc-fcs}) presents itself a nice consistency check.
In $\cN=1$ superconformal field theory,
the three-point function $\langle J_{\a \ad}\,J_{\b \bd}\, L\rangle$
of two supercurrents with one flavour current superfield $L$
is uniquely determined up to an overall constant \cite{osborn}.
Any $\cN=2$ superconformal field theory, considered as a 
particular $\cN=1$ superconformal model, possesses 
a special flavour current superfield, $L=J$.
Therefore, the only possible arbitrariness in the
structure of the correlation function 
$\langle J_{\a \ad}\,J_{\b \bd}\, J\rangle$
is an overall constant.    
But $J$ and $J_{\a \ad}$ are parts of
the $\cN=2$ supercurrent $\cJ$, and hence
the three-point function
$\langle J_{\a \ad}\,J_{\b \bd}\, J \rangle$ follows
from $\langle \cJ\, \cJ\, \cJ \rangle$. Since the latter 
contains two linearly independent forms, 
given in eq. (\ref{sc3-pf}), there are two 
possibilities: 
(i) either $A$-term or $B$-term in 
(\ref{sc3-pf}) does not contribute to 
$\langle J_{\a \ad}\,J_{\b \bd}\, J \rangle$; 
(ii) both $A$-term and $B$-term
produce the same functional contribution 
to $\langle J_{\a \ad}\,J_{\b \bd}\, J \rangle$ modulo
overall constants. Eq. (\ref{sc-sc-fcs}) 
tells us that option (ii) is realized.

The calculation of $\langle J_{\a\dot\alpha}J_{\beta\dot\beta}
J_{\gamma\dot\gamma}\rangle$ is much more tedious. To derive 
this correlation function, one has to act with the operator
(\ref{scoperator}) on each argument of 
$\langle \cJ (z_1) \cJ (z_2) \cJ (z_3) \rangle$. 
But since by construction $H$ in (\ref{sc3-pf0}) is 
a function of $\bZ_3$, it turns out to be quite difficult
to control superconformal covariance at 
intermediate stages of the calculation when acting 
with $\D_{ \g \dot{\g} }$ on the third argument of 
$\langle \cJ (z_1) \cJ (z_2) \cJ (z_3) \rangle$.   
A way out is as follows. One first computes
\bea
&& \langle \D_{\a \ad} \cJ (z_1) \cJ (z_2) \cJ (z_3) \rangle
= 
\frac{
(\tilde{x}_{1 \bar{3}}{}^{-1})_{\a \dot{\s}}
(\tilde{x}_{3 \bar{1}}{}^{-1})_{\s \dot{\a}}
}{ 
x_{\bar{1}3}{}^2 x_{\bar{3}1}{}^2 
x_{\bar{2}3}{}^2 x_{\bar{3}2}{}^2} \non \\
& \times & \Bigg( 
\frac{64}{3}\,
\q^\s_{13 \,\underline{2}}\,
{\bar \q}_{13}^{\dot{\s} \,\underline{2} } \; H (\bZ_3)
+ \frac{16}{3}\, \q^\s_{13\,\underline{2} }\;
u_k{}^{ \underline{2} }(z_{31}) \;
{\bar \cD}^{\dot{\s} k} \, H (\bZ_3)
- \frac{16}{3} \,
{\bar \q}_{13}^{\dot{\s} \,\underline{2} }\; 
u_{\underline{2} }{}^k(z_{13})\; \cD^\s_k \,H (\bZ_3) \non \\
&+& \hf \,\Big\{ u_{\underline{2} }{}^k(z_{13})\;
u_l{}^{ \underline{2} }(z_{31})
-\frac{1}{3}\, u_{\underline{1} }{}^k (z_{13})\;
u_l{}^{ \underline{1} }(z_{31}) \Big\}\;
[ \cD^\s_k \, , \, {\bar \cD}^{\dot{\s} l}]\, H (\bZ_3)  
\Bigg)
\non
\eea
and next expresses $H (\bZ_3)$, $\cD^\s_k \,H (\bZ_3)$
and $[ \cD^\s_k  ,  {\bar \cD}^{\dot{\s} l}]\, H (\bZ_3)$
as functions of $\bZ_1$ with the help of identities (\ref{difz}).
After that it is a simple, but time-consuming procedure,
to complete the computation of 
$\langle J_{\a\dot\alpha}J_{\beta\dot\beta}
J_{\gamma\dot\gamma}\rangle$. The result reads
\bea
\langle
J_{\a \ad} (z_1)\, J_{\b \bd} (z_2)\, J_{\g \dot{\g}} (z_3)
\rangle & = &
\frac{
(x_{1 \bar{3}})_{\a \dot{\s}}
(x_{3 \bar{1}})_{\s \dot{\a}}
(x_{2 \bar{3}})_{\b \dot{\d}}
(x_{3 \bar{2}})_{\d \dot{\b}}
}{ 
(x_{\bar{1}3}{}^2 x_{\bar{3}1}{}^2 
x_{\bar{2}3}{}^2 x_{\bar{3}2}{}^2)^2 }\;
H^{\dot{\s} \s, \dot{\d} \d}{}_{\g \dot{\g}} 
(\bX_3, \bar \bX_3)~, 
\non \\
H^{\dot{\s} \s, \dot{\d} \d }{}_{\g \dot{\g}} 
(\bX_3, \bar \bX_3) & = & 
h^{\dot{\s} \s, \dot{\d} \d}{}_{\g \dot{\g}} 
(\bX_3, \bar \bX_3) + 
h^{ \dot{\d} \d, \dot{\s} \s }{}_{\g \dot{\g}} 
(- \bar \bX_3, - \bX_3) ~,
\label{final1}
\eea
where 
\bea
h^{\ad \a, \bd \b}{}_{\g \dot{\g}} 
(\bX, \bar \bX) &=& \frac{64}{27} \,(26A-\frac{9}{4}B) 
\; \frac{ {\rm i}  }  
{(\bX^2)^2}\; \bX^{\bd \a}  \d^\ad_{\dot \g} \d^\b_{{\g}} \non  \\
&-& \frac{8}{27}\, (8A-9B)\; \frac{ {\rm 1}  }  
{(\bX^2)^3}\;
\Bigg( 2\Big(\bX^{\ad \a} \bP^{\bd \b} 
+ \bX^{\bd \b} \bP^{\ad \a} \Big) \bX_{\g \dot{\g}} \non \\
&-& 3 \bX^{\ad \a} \bX^{\bd \b}\Big(\bP_{\g \dot{\g}}
+2 \frac{(\bP \cdot \bX)}{\bX^2} \bX_{\g \dot{\g}}\Big)  \\
&+& 2\Big( (\bP \cdot \bX) \bX^{\a \ad} -\bX^2 \bP^{\a \ad}\Big)
\d_\g^\b \d^\bd_{\dot{\g}} 
+  2\Big( (\bP \cdot \bX) \bX^{\b \bd} -\bX^2 \bP^{\b \bd}\Big)
\d_\g^\a \d^\ad_{\dot{\g}} \non \\
&+& \Big(4(\bP \cdot \bX) \bX^{\a \bd} +\bX^2 \bP^{\a \bd}\Big)
\d_\g^\b \d^\ad_{\dot{\g}}
+ \Big(4(\bP \cdot \bX) \bX^{\b \ad} +\bX^2 \bP^{\b \ad}\Big)
\d_\g^\a \d^\bd_{\dot{\g}} \Bigg)~.\non
\eea
It is convenient to rewrite this result in vector notation
\bea
h^{abc}(\bX, \bar \bX) & \equiv & 
-\frac{1}{8}\,(\s^a)_{\a \ad}\,(\s^a)_{\b \bd}\,
(\tilde{\s}^c)^{\dot{\g} \g}\; h^{\ad \a, \bd \b}{}_{\g \dot{\g}} 
(\bX, \bar \bX) 
\label{final2}\\
&=& - \frac{16}{27} \,(26A-\frac{9}{4}B) 
\; \frac{ {\rm i}  } {(\bX^2)^2}\;
\Big(\bX^a \eta^{bc} + \bX^b \eta^{ac} - \bX^c \eta^{ab} 
+{\rm i}\, \ve^{abcd} \bX_d \Big) \non \\
&&- \frac{8}{27}\, (8A-9B)\; \frac{ {\rm 1}  }  
{(\bX^2)^3}\;\Bigg( 
2\Big(\bX^{a} \bP^{b} 
+ \bX^{b} \bP^{a} \Big) \bX^c -3\bX^a \bX^b \Big( \bP^c
+2\frac{(\bP \cdot \bX)}{\bX^2} \bX^c \Big) \non  \\
&& - (\bP \cdot \bX)\Big( 
3( \bX^a \eta^{bc} + \bX^b \eta^{ac} ) -2 \bX^c \eta^{ab} \Big)
+\hf \bX^2 \Big( \bP^a \eta^{bc} + \bP^b \eta^{ac} 
 + \bP^c \eta^{ab} \Big) \Bigg)~. \non
\eea 
Our final relations (\ref{final1}) and (\ref{final2})
perfectly agree with the general structure of the three-point
function of the supercurrent in $\cN=1$ superconformal 
field theory \cite{osborn}.

Using the results of \cite{osborn}, it is easy to express
$A$ and $B$ in terms of the anomaly coefficients \cite{Anselmietal} 
\be
a = \frac{1}{24}(5n_V +n_H)~, \qquad \quad c= \frac{1}{12}(2n_V +n_H)~,
\ee 
where $n_V$ and $n_H$ denote the number of free $\cN = 2$
vector multiplets and hypermultiplets, respectivley.
We obtain\footnote{Our definition of the $\cN=1$ supercurrent
corresponds to that adopted in \cite{bk} and differs in sign
from Osborn's convention \cite{osborn}.}  
\be
A= \frac{3}{64 \p^6}(4a - 3c)~, \qquad \quad
B= \frac{1}{8 \p^6}(4a - 5c)~.
\ee

In $\cN=1$ supersymmetry, a superconformal Ward identity
relates the coefficient in the two-point function of the 
supercurrent (\ref{n=1sct-pf}) to the anomaly coefficient
$c$ as follows \cite{osborn}
\be
c_\cJ = \frac{3}{8\p^4} \,c~.
\ee
In terms of the coefficients $A$ and $B$ this relation reads
\be
\frac{2}{\p^2}\,c_\cJ =  8A-3B ~.
\ee

In $\cN=1$ supersymmetry, there also exists a superconformal Ward identity
which relates the coefficients in the following correlation functions
\bea
\langle
L (z_1) \, L (z_2)
\rangle  &=&
\frac{c_L }
{  x_{ {\bar 1} 2}{}^2  x_{{\bar 2} 1}{}^2} ~,\non \\
\langle
L (z_1)\, L (z_2) \,J_{\a \ad} (z_3) 
\rangle & = &
\frac{D}{ 
x_{\bar{1}3}{}^2 x_{\bar{3}1}{}^2 
x_{\bar{2}3}{}^2 x_{\bar{3}2}{}^2 } 
%\non \\
%& \times & 
\left( 
\frac{ 2 (\bP_3 \cdot \bX_3) \bX_{3\, \a \ad} 
+ \bX_3{}^2 \bP_{3\, \a \ad}  }
{(\bX_3{}^2)^2 } 
~+~ {\rm c.c.}
%(\bX_3  \leftrightarrow  - \bar{\bX}_3)
%\frac{ \bX_{3\, \a \ad} }{ \bar{\bX}_3{}^2 }
%- \frac{ \bar{\bX}_{3\, \a \ad} }{\bX_3{}^2}
\right) \non 
\eea
of a current superfield $L$. A nice consequence
of our consideration is that 
$\cN=2$ supersymmetry allows us to fix up this
Ward identity without working it out explicitly.
The point is that the $\cN=2$ supercurrent contains a
special current superfield, that is $J$. Therefore, from 
the first relation in (\ref{n=1sct-pf}) and eq. (\ref{JJ-sc})
we deduce
\be
D= -\frac{1}{6\p^2}\,c_L ~.
\label{wi}
\ee

${}$Let us turn to the three-point function 
of the $\cN=2$ flavour current superfield given by eqs.
(\ref{abc}) and (\ref{n=2fcs3p-f2}). 
${}$From these relations one reads off
the three-point function of the $\cN=1$ component
(\ref{n=1fcs}) 
\be
\langle
L^{\overline{a}} (z_1)\, L^{\overline{b}} (z_2)\, 
L^{\overline{c}} (z_3)
\rangle  ~=~ 
\frac{1}{4}\,f^{\overline{a} \overline{b} \overline{c}} 
\frac{{\rm i}}{ 
x_{\bar{1}3}{}^2 x_{\bar{3}1}{}^2 
x_{\bar{2}3}{}^2 x_{\bar{3}2}{}^2}
\left(  \frac{1}{{\bar \bX}_3{}^2 }
- \frac{1}{ \bX_3{}^2}
\right) ~.
\label{n=2t-pffcs2}
\ee
Here we have used the identities
\be
{\bf u}_{\underline{1}}{}^{\underline{1}}(\bZ_3)|
= \det {\bf u}(\bZ_3) |~,
\qquad {\bf u}_{\underline{1}}{}^{\underline{2}}(\bZ_3) |=
{\bf u}_{\underline{2}}{}^{\underline{1}}(\bZ_3) | =0~,\qquad
{\bf u}_{\underline{2}}{}^{\underline{2}}(\bZ_3) | =1~.
\ee
It is worth noting that Ward identities allow to represent
$f^{\overline{a} \overline{b} \overline{c}}$
as a product of $c_\cL$ and the structure constants of
the flavour symmetry group, see \cite{osborn} for more details.

In $\cN=1$ superconformal 
field theory, the three-point 
function of flavour current superfields $L$
contains, in general, two linearly independent forms \cite{osborn}:
$$
\langle
L^{\overline{a}} (z_1)\, L^{\overline{b}} (z_2)\, 
L^{\overline{c}} (z_3)
\rangle  = 
\frac{1}{ 
x_{\bar{1}3}{}^2 x_{\bar{3}1}{}^2 
x_{\bar{2}3}{}^2 x_{\bar{3}2}{}^2} 
\Bigg\{
{\rm i}\, f^{[\overline{a} \overline{b} \overline{c}]} 
\left(\frac{1}{ \bX_3{}^2}-  
\frac{1}{{\bar \bX}_3{}^2 } \right) 
+  d^{(\overline{a} \overline{b} \overline{c})} 
\left(\frac{1}{ \bX_3{}^2}+  
\frac{1}{{\bar \bX}_3{}^2 } \right)\Bigg\}~.
$$ 
The second term, involving a completely symmetric
group tensor $d^{\overline{a} \overline{b} \overline{c}}$,
reflects the presence of chiral anomalies in the theory. 
The field-theoretic origin of this term is due to the
fact that the $\cN=1$ conservation equation
$\bar{D}^2\, L = D^2\, L = 0$ admits a non-trivial deformation
\bea
\bar{D}^2\, \langle L^{ \overline{a} } \rangle ~\propto ~
d^{\overline{a} \overline{b} \overline{c}}\;
W^{\overline{b}\, \a} \, W^{\overline{c}}_\a \non
\eea
when the chiral flavour current is coupled 
to a background vector multiplet.  
Eq. (\ref{n=2t-pffcs2}) tells us that the
flavour currents are anomaly-free in $\cN=2$
superconformal theory. This agrees 
with the facts that (i) $\cN=2$ super Yang-Mills
models are non-chiral; (ii)
the $\cN=2$ conservation
equation (\ref{fcce}) does not possess 
non-trivial deformations.
%The structure of the correlation function obtained  
%is compatible with $\cN=1$ superconformal considerations
%\cite{osborn}.

{}Finally, from the three-point function (\ref{l-l-j}) 
we immediately deduce
\bea
\langle
L^{\overline{a}} (z_1)\, L^{\overline{b}} (z_2)\, 
J (z_3)
\rangle  &=& 
\frac{d}{2}\;\d^{\overline{a} \overline{b}} \; 
\frac{{\rm 1}}{ 
x_{\bar{1}3}{}^2 x_{\bar{3}1}{}^2 
x_{\bar{2}3}{}^2 x_{\bar{3}2}{}^2}
\left(  \frac{1}{ \bX_3{}^2}
+ \frac{1}{{\bar \bX}_3{}^2 }
\right) ~, \non \\
\langle L^{\overline{a}} (z_1)\, L^{\overline{b}} (z_2)\, 
J_{\a \ad} (z_3)
\rangle  &=& 
-\frac{2d}{3}\;\d^{\overline{a} \overline{b}} \;
 \frac{1}{ 
x_{\bar{1}3}{}^2 x_{\bar{3}1}{}^2 
x_{\bar{2}3}{}^2 x_{\bar{3}2}{}^2 }  \\
& \times &
 \left( 
\frac{ 2 (\bP_3 \cdot \bX_3) \bX_{3\, \a \ad} 
+ \bX_3{}^2 \bP_{3\, \a \ad}  }
{(\bX_3{}^2)^2 } 
~+~(\bX_3  \leftrightarrow  - \bar{\bX}_3)
\right)~. \non
\eea
Now, the Ward identity (\ref{wi}) implies
\be
d ~=~ \frac{1}{4\p^2}\; c_\cL~.
\ee

\section{Discussion}
\setcounter{footnote}{0}
Our main objective in this paper was to determine the restrictions 
of the general structure of two-- and three-- point functions of 
conserved currents imposed by $\cN=2$ superconformal symmetry. 
This was done in a manifestly supersymmetric formalism. The results
are contained in sects. 3.2 and 3.3. In particular, we have shown that 
the three--point function of the supercurrent allows for two 
independent structures.
In the appendices we show that the minimal supergravity
multiplet can be described in harmonic superspace 
by two real unconstrained prepotentials: harmonic $G$ and 
analytic $v^{++}_5$. This is the superfield parametrisation 
which allows us to derive the supercurrent and multiplet 
of anomalies as the response of the matter action to small 
disturbances of the supergravity prepotentials.

In this paper, the results about the structure of the 
correlation functions were completely determined  
by $\cN=2$ superconformal symmetry. The results for specific models 
only differ in the value of the numerical coefficients. 
They can be determined in perturbation theory using 
supergraph techniques. 

An interesting open problem is the issue of non-renormalization 
theorems for the correlation functions of conserved currents. 
For a recent discussion for $\cN=4$, see \cite{PS}.

There exists an off-shell formulation of $\cN=3$ SYM theory, 
ref. \cite{GIKOS3}.
Since $\cN=3$ and $\cN=4$ SYM are dynamically equivalent, 
it can be used to get further restrictions and possible 
non-renormalization theorems on the $\cN=4$ correlation functions. 

Another interesting problem is the structure of superconformal anomalies
of $\cN=2$ matter systems in a supergravity background. Such anomalies 
are responsible for the three-point function of the $\cN=2$ supercurrent 
studied in sect. 3. The results of Apps. A and B provide the natural 
prerequisites for the analysis of the $\cN=2$ superconformal anomalies.

\vskip.5cm

\noindent
{\bf Acknowledgement} \hfill\break
We would like to thank H. Osborn for a helpful email correspondence, 
J. Gates for pointing out relevant references and 
G. Arutyunov and J. Pawelczyk for useful discussions. This work was 
supported in parts by the DFG-SFB-375 grant, by GIF--the German-Israeli 
Foundation for Scientific Research, by the European Commission TMR
programme ERBFMRX-CT96-0045 and by the NATO grant PST.CLG 974965,
DFG--RFBR grant 96-02-00180-ext., RFBR grant 99-02-16617, INTAS grant 96-0308.

\begin{appendix}

\section{Supergravity multiplets}
\setcounter{footnote}{0}
In this appendix we briefly review harmonic superspace and 
discuss the Weyl multiplet and the minimal supergravity multiplet 
in some detail.

In rigid supersymmetry, all known $\cN=2$ supersymmetric theories
in four space-time dimensions can be described in terms of fields
living in $\cN=2$ harmonic superspace ${\Bbb R}^{4|8} \times 
{\rm SU(2)/U(1)}$ introduced by GIKOS \cite{gikos}.
Along with the standard coordinates 
$z = (x^m, \q^\a_i, {\bar \q}_\ad^i )$ of ${\Bbb R}^{4|8}$ 
(${\bar \q}^{\ad i} = \overline{\q^\a_i}$), this superspace involves
the internal harmonic variables $u^\pm_i$ which are constrained by 
$u^{+i} u^-_i = 1$ and defined modulo phase rotations with 
charge $\pm 1$. Harmonic superspace possesses a supersymmetric 
subspace , with half the fermionic coordinates of the full superspace,
defined to be spanned by the variables
\be
\left\{ \z^M , u^\pm_i \right\}~, \qquad \quad
\z^M = (x^m_A, \q^{+ \hat{\a}}) = 
 (x^m_A, \q^{+ \a}, {\bar \q}^{+ \ad})
\label{analsub}
\ee
where \footnote{Eq. (\ref{analbasis}) defines the so-called analytic basis
of harmonic superspace, while the original basis $\{ z, u^\pm_i \}$
is called central. In what follows, we mainly use the analytic basis 
and do not indicate the subscript `$A$' explicitly.}
 \be
x^m_A = x^m - 2{\rm i} \q^{(i} \s^m {\bar \q}^{j)}
u^+_i u^+_j ~, \qquad \quad 
\q^{\pm \hat{\a}} = \q^{\hat{\a} i} u^\pm_i \;.
\label{analbasis}
\ee
Analytic subspace (\ref{analsub}) is closed
under $\cN=2$ super Poincar\'e 
and superconformal transformations \cite{gikos,gios1}.
In addition, it is invariant under the generalized conjugation 
` $\breve{}$ ' defined as \cite{gikos}:
$$
\breve{}~:~ ~x^m_A \rightarrow x^m_A~, \quad
\q^+_\a \rightarrow {\bar \q}^+_\ad ~, \quad 
{\bar \q}^+_\ad \rightarrow -\q^+_\a~, \quad 
u^{\pm i} \rightarrow - u^\pm_i~, \quad
  u^\pm_i  \rightarrow u^{\pm i}\;.
$$
The fundamental importance of analytic subspace (\ref{analsub})
lies in the fact that the $\cN=2$ matter multiplets (hypermultiplets and 
vector multiplets) can be described in terms of unconstrained 
analytic superfields living in the analytic subspace (\ref{analsub}).

In harmonic superspace, there is a universal gauge principle
to introduce couplings to Yang-Mills and supergravity \cite{gikos}. 
Consider the rigid supersymmetric operators $D^{++}$ and $D^{--}$
defined as
\be
D^{\pm \pm} = \pa^{\pm \pm} - 2{\rm i} \q^\pm \s^m {\bar \q}^\pm  \pa_m
 + \q^{\pm \hat{\a} } \pa^\pm _{\hat{\a}} ~,
\label{greatd}
\ee
where $\pa^{\pm \pm} = u^{\pm i} \pa / \pa u^{\mp i}  ~,~ 
\pa_m = \pa / \pa  x^m_A  ~,~
\pa^\pm_{\hat{\a} }  = \pa / \pa \q^{\mp \hat{\a} }  $.
The fundamental  property of $D^{++}$ is that if $\phi$ is analytic, i.e. if 
$\partial^+_{\hat\alpha}\phi=0$, then so is $D^{++}\phi$.
It turns our that switching on the  Yang-Mills or 
supergravity couplings is equivalent to  
the requirement that $D^{++}$ must be deformed to acquire a connection
or nontrivial vielbeins, in such a way that the deformed 
operator still preserves analyticity.

\subsection{Weyl multiplet}

In this subsection we start with reviewing the harmonic superspace
realization \cite{gios1,gios2} of the Weyl multiplet \cite{dwsugra}
describing $N=2$ conformal supergravity and 
comprising $24 + 24$ off-shell degrees of freedom. Then, we will
present a new parametrisation for  the conformal supergravity
prepotentials and describe several gauge fixings.

%The Weyl multiplet \cite{dwsugra}, which describes $\cN=2$ conformal 
%supergravity and comprises $24 + 24$ off-shell degrees of freedom, 
%can be realized in harmonic superspace
%\cite{gios1,gios2}.
%After a brief review of these facts we will
%present a new parametrisation for the conformal supergravity
%prepotentials and also describe several gauge fixings.

According to \cite{gios1,gios2},
the conformal supergravity gauge fields are identified with the vielbein 
components of a real covariant derivative
\be 
\cD^{++} = \pa^{++} + \cH^{++M} \pa_M + \cH^{(+4)} \pa^{--} 
+\cH^{+\hat{\a}} \pa^+_{\hat{\a}}
\label{covd1}
\ee
that is required to move every analytic superfield into an analytic
one. Hence $\cH^{++M} \equiv ( \cH^{++m}, \cH^{+++\a},  \breve{\cH}^{+++\ad})$
and $\cH^{(+4)}$ are analytic, while $\cH^{+\hat{\a}} \equiv (\cH^{+\a}, 
\breve{\cH}^{+\ad})$ are unconstrained superfields. The supergravity 
gauge transformations act on $\cD^{++}$ and a scalar superfield $U$
via the rule ($D^0=u^{+i}{\partial\over\partial{u^{+i}}}-
u^{-i}{\partial\over\partial {u^{-i}}}$)
\be
\d \cD^{++} = [\l + \r , \cD^{++}] + \l^{++} D^0~, \qquad \qquad
\d U = (\l + \r) U
\label{trlaw1}
\ee
where 
\be
\l = \l^M \pa_M + \l^{++} \pa^{--}~,
\qquad \quad \r = \r^{- \hat{\a}} \pa^+_{\hat{\a}}
\label{par1}
\ee
such that every analytic superfield of U(1) charge $p$,  
$\F^{(p)}$, remains analytic
\be
\d \F^{(p)} = \l \F^{(p)} ~, \qquad  \quad \pa^+_{\hat{\a}}\, \F^{(p)} 
= \pa^+_{\hat{\a}}  \,\d \F^{(p)} = 0\;.
\ee
Therefore, the parameters $\l^M = (\l^m, \l^{+\a}, \breve{\l}^{+\ad} )$
and $\l^{++}$ are analytic, while $\r^{- \hat{\a}} = (\r^{-\a}, 
\breve{\r}^{-\ad})$ are unconstrained superfields.

The supergravity gauge transformations are induced by 
special reparametrisations of  harmonic superspace
\bea
\d \z^M &=& - \l^M(\z,u)~, \non \\
\d u^{+i} &=& - \l^{++}(\z,u) u^{-i}~, 
\quad \qquad \d u^{-i} =  0~, \non \\
\d \q^{- \hat{\a}} &=& -\r^{-\hat{\a}} (\z, \q^-, u)
\label{gctr}
\eea
which leave the analytic subspace invariant.

Since $\cD^{++}$ contains a number of independent vielbeins,
it is far from obvious in the above picture how to generate
a single scalar supercurrent from the host of harmonic vielbeins.
In addition,  there is a technical complication -- some vielbeins
possess non-vanishing values in the flat superspace limit
(\ref{greatd}). To find a way out, it is sufficient to recall
the standard wisdom of superfield $\cN=1$ supergravity \cite{sg}. 
In eqs. (\ref{covd1}) and (\ref{par1}) the covariant derivative and gauge
parameters are decomposed with respect to the superspace partial 
derivatives. To have a simple flat superspace limit (which would
correspond to vanishing values for all the supergravity prepotential),
it is convenient to decompose $\cD^{++}$ and $\l,\; \r$ with
respect to flat covariant derivatives $D^{\pm \pm}$,
$D_M = (\pa_m, D^-_\a, {\bar D}^-_{\ad})$ and 
$D^+_{\hat{\a}} = \pa / \pa\q^{-\hat{\a}}$; i.e.
\bea
\cD^{++} &=& D^{++} + H^{++M}D_M +H^{(+4)} D^{--}
+ H^{+\hat{\a}} D^+_{\hat{\a}}
\label{param} \\
\l &=& \L^M D_M + \L^{++} D^{--} \qquad \quad 
\r = \r^{-\hat{\a}} D^+_{\hat{\a}} 
\label{par2}
\eea
where $H^{(+4)} = \cH^{(+4)}$, $\L^{++} = \l^{++}$.
In such a parametrisation, the vielbeins $H^{++M}$ and 
and $H^{(+4)}$ are no longer independent, but they are instead 
expressed via a single unconstrained superfield.
Really, since we must have 
$$
D^+_\a \cD^{++} \F^{(p)} = 0 ~,
$$
for any analytic superfield $\F^{(p)}$,
using the algebra of flat covariant 
derivatives leads to
\bea
& D^+_\a H^{++\b \bd} - 2{\rm i} \d_\a^\b 
\breve{H}^{+++\bd} = 0 ~, \qquad \quad D^+_\a H^{(+4)} = 0 ~, \non \\
& D^+_\a  H^{+++\b} - \d_\a^\b H^{(+4)} = 0 ~, \qquad \quad
D^+_\a \breve{H}^{+++\bd} = 0\;.
\label{prepconstr1}
\eea
The general solution to these equations (and their conjugates) reads
\bea
& H^{++\a \ad} = - {\rm i} \, D^{+ \a} {\bar D}^{+ \ad} G ~, \non \\
& H^{+++\a} = - \frac{1}{8} D^{+\a} ({\bar D}^+)^2 G~, \qquad 
\breve{H}{}^{+++\ad} =  \frac{1}{8} {\bar D}^{+\ad}
(D^+)^2 G ~, \non \\
& H^{(+4)} = \frac{1}{16} (D^+)^2 ({\bar D}^+)^2 G
\equiv  (D^+)^4 G
\label{prepot1}
\eea
with $G(\z, \q^-, u)$ a real unconstrained harmonic 
superfield, $\breve{G} = G$.
The prepotential introduced is defined modulo 
pre-gauge transformations
\be
\d G = \frac{1}{4} (D^+)^2 \O^{--} + \frac{1}{4}
({\bar D}^+)^2 \breve{\O}^{--}
\label{pre-gauge}
\ee
where $\O^{--}$ is a complex unconstrained parameter.

Similar to $H^{++M}$ and $H^{(+4)}$, the gauge parameters 
$\L^M$ and $\L^{++}$ in eq. (\ref{par2}) are expressed 
via a single real unconstrained superfield
$l^{--} (\z, \q^-, u)$, $\breve{l}^{--} = l^{--}$, 
as 
\bea 
& \L^{\a \ad} = - {\rm i} \, D^{+ \a} {\bar D}^{+ \ad} l^{--} ~,
\qquad  \quad \L^{++} =   (D^+)^4 l^{--}~,
\non \\
& \L^{+\a} = -\frac{1}{8} D^{+\a} ({\bar D}^+)^2 l^{--}~, \qquad 
\breve{\L}{}^{+\ad} =  \frac{1}{8} {\bar D}^{+\ad}
(D^+)^2 l^{--} \;.
\label{prepot2}
\eea

${}$From eq. (\ref{trlaw1}) one can read  off the transformations
of $H^{++M}$, $H^{(+4)}$ and $H^{+\hat{\a}}$ :
\bea 
& \d H^{++M} =\l \; H^{++M} - \tilde{\cD}^{++} \L^M ~,
\qquad \quad  
\d H^{(+4)} =\l \; H^{(+4)} - \cD^{++} \L^{++} ~, \non \\
& \d H^{+ \hat{\a}} = (\l + \r) H^{+ \hat{\a}} - \L^{+\hat{\a}}
- \cD^{++} \r^{-\hat{\a}}~,
\label{trlaw2}
\eea
where 
$$
\tilde{\cD}^{++} = \cD^{++} - H^{+ \hat{\a}} D^+_{\hat{\a}} \;.
$$
Since the parameters $\r^{-\hat{\a}}$ are unconstrained, 
$H^{+ \hat{\a}}$ can be gauged away
\be
H^{+ \hat{\a}} = 0\;.
\label{goodgauge}
\ee
Then, the residual gauge freedom is constrained by
\be
\cD^{++} \r^{- \hat{\a}} = - \L^{+ \hat{\a}}\;.
\ee
In what follows, we will assume eq. (\ref{goodgauge}), 
hence $\cD^{++}$ and $\tilde{\cD}^{++}$ coincide.

${}$From (\ref{trlaw2}) it is easy to read off the transformation 
law of $G$. It is sufficient to notice the identities
\be
[D^+_{\hat{\a}} , \l] = 0 ~, \qquad \quad 
[D^+_{\hat{\a}} , \cD^{++}] = 0
\ee
where the latter holds for (\ref{goodgauge}) only. Therefore,   
from eqs. (\ref{prepot1}), (\ref{prepot2}) and  (\ref{trlaw2})
we deduce
\be 
\d G = \l G - \cD^{++} l^{--}\;.
\label{Gtrlaw}
\ee
Now, eqs. (\ref{pre-gauge}) and (\ref{Gtrlaw})
determine the full supergravity gauge group.

It is instructive to examine (\ref{Gtrlaw}) in linearized theory
\be 
\d G = - D^{++} l^{--} \;.
\label{linearized}
\ee
In the central basis $D^{++}$ coincides with $\pa^{++}$,
and $G(z,u)$ and $l^{--} (z,u)$ are
\bea
G(z,u) &=& \bG(z) + \sum_{n=1}^{\infty} 
G^{(i_1 \cdots i_n j_1\cdots j_n)} (z)
u^+_{i_1} \cdots u^+_{i_n} 
u^-_{j_1} \cdots u^-_{j_n} ~, \non \\
l^{--}(z,u) &=& \sum_{n=1}^{\infty} 
l^{(i_1 \cdots i_{n-1} j_1\cdots j_{n+1})} (z)
u^+_{i_1} \cdots u^+_{i_{n-1}} 
u^-_{j_1} \cdots u^-_{j_{n+1}}
\eea
where $\bG(z)$, $G^{(i_1 \cdots i_{2n})}(z)$ and 
$l^{(i_1 \cdots i_{2n})}(z)$ are unconstrained 
superfields. Since $D^{++} u^+_i = 0$ and 
$D^{++} u^-_i = u^+_i$, eq. (\ref{linearized})
tells us that all the components
$G^{(i_1 \cdots i_{2n})}$, $n = 1,2, \dots$, 
can be gauged away to arrive at the gauge
condition
\be
D^{++} G = 0\;.
\label{firstgc}
\ee
The surviving gauge freedom consists of those 
combined transformations 
(\ref{pre-gauge}) and (\ref{Gtrlaw}) which 
preserve the above gauge condition, that is
\be
 \d \bG(z) = \frac{1}{12} D_{ij} \O^{ij}(z) + 
\frac{1}{12} {\bar D}_{ij} {\bar \O}^{ij}(z) 
\label{lingfr}
\ee
where $\O^{ij}(z)$ is the leading component in the harmonic
expansion of $\O^{--}(z,u)$  (\ref{pre-gauge}).
The linearized prepotential of conformal supergravity $\bG(z)$ and 
its gauge freedom (\ref{lingfr}) is precisely what follows from the 
structure of the $\cN=2$ supercurrent discussed in the introduction.

Instead of imposing the gauge condition (\ref{firstgc}), one can 
take a different course. Since $H^{(+4)}$ is analytic, 
it follows from (\ref{trlaw2}) that 
we can achieve the gauge \cite{gios1,gios2}
\be 
H^{(+4)} = 0
\label{badgauge}
\ee
which restricts the residual gauge freedom to
\be
\cD^{++} \L^{++} = 0\;.
\label{badgaugepar}
\ee
Now, from (\ref{prepot1}) and (\ref{badgauge}) we get
\be
G = D^{+ \a} \J^-_\a  ~+~ {\bar D}^{+ \ad} 
\breve{\J}^-_\ad
\ee
where $\J^-_\a (z,u)$ is an unconstrained harmonic spinor superfield
of U(1) charge $-1$. Eq.  (\ref{badgaugepar}) defines a linear analytic 
superfield in conformal supergravity background. In {\it linearized} theory,
the general solution of eq. (\ref{badgaugepar}) is well known:
\be
\L^{++} = (D^+)^4 \Big\{ u^-_i u^-_j    D^{ij} V(z) 
+u^-_i u^-_j    {\bar D}^{ij} \bar V (z) \Big\} \;. 
\ee
Therefore, from here and (\ref{prepot2}) we can completely 
specify the residual gauge freedom:
\be
l^{--} (z,u) = 
D^{+ \a} \U^{(-3)}_\a (z,u)  + {\bar D}^{+ \ad} 
\breve{\U}^{(-3)}_\ad (z,u)  ~+~
u^-_i u^-_j  \Big(  D^{ij} V(z) 
+   {\bar D}^{ij} \bar V (z) \Big)
\ee
with an unconstrained harmonic parameter $ \U^{(-3)}_\a (z,u)$.
Using $\U^{(-3)}_\a$ transformations, we can gauge away
all  $\J^-_\a $ but the leading component in its harmonic expansion
\be
\J^-_\a (z,u) = \J^i_\a (z) u^-_i \;.
\ee
$ \J^i_\a (z)$ is nothing but the Gates-Siegel 
prepotential \cite{gs}.

\subsection{Minimal multiplet}

The so-called minimal supergravity multiplet \cite{dwsugra}
is obtained by coupling the Weyl multiplet to 
an Abelian  vector multiplet which is 
a real analytic superfield $\cV^{++}_5 (\z,u)$.
$\cV^{++}_5 (\z,u)$
transforms as a scalar under  (\ref{trlaw1}) and 
possesses its own gauge freedom \cite{gikos,gios1,gios2}
\be
\d \cV^{++}_5 = - \cD^{++} \l_5
\ee
with $\l_5 (\z,u)$ an arbitrary  real analytic parameter. 
This vector multiplet is a gauge
field for the central charge $\D$ that can be understood as the 
derivative in an extra bosonic coordinate $x^5$, 
$\D = \pa / \pa x^5$, on which matter supermultiplets may 
depend. For matter supermultiplets with central 
charge, the definition (\ref{covd1}) should be replaced by
\be
\cD^{++}_\D = \cD^{++} + \cV^{++}_5 \D
\label{covd2}
\ee
and the transformations (\ref{trlaw1}) extend to
\be
\d \cD^{++}_\D = [\l + \r  +\l_5 \D \, , \, \cD^{++}_\D] + \l^{++} D^0 ~,
\qquad \qquad
\d U = (\l + \r  + \l_5 \D) U\;.
\label{trlaw3}
\ee

The limit of rigid supersymmetry corresponds to the 
choice when all $H$--vielbeins
in (\ref{param}) vanish and $\cV^{++}_5$ can be brought to the form 
\be
\cV^{++}_{5,{\rm flat}} = {\rm i}\; 
\left( (\q^+)^2 - ({\bar \q}^+)^2 \right)\;.
\label{rigidcc}
\ee
That is why $\cV^{++}_5$ must in general satisfy a global 
restriction that its scalar component field $\cZ(x) $ defined by 
\bea
\cV^{++}_5 (\z, u) & \sim & (\q^+)^2 \breve{ \cZ} (x,u) + 
({\bar \q}^+)^2 \cZ (x,u) ~, \non \\
\cZ (x,u) & = & \cZ (x)  + \sum_{n=1}^{\infty}
\cZ^{ 
( i_1 \cdots i_n  j_1 \cdots j_n ) 
} (x)
u^+_{{i_1}} \cdots u^{+}_{i_n} 
{}
u^{-}_{j_1} \cdots u^-_{j_n}  
\label{decomposi}
\eea
be non-vanishing over the space-time,
$ \cZ (x) \neq 0 $.
Then, ordinary local scale and chiral 
transformations (contained in (\ref{par2})) 
can be used to bring $\cZ (x)$ to its flat 
form (\ref{rigidcc}) (all remaining components
in (\ref{decomposi}) turn out to be  
gauge degrees of freedom). Let $\bD_\a^i$, 
$\bar{\bD}_{\ad i}$ 
and $\bD^{\pm \pm}$ be the flat covariant derivatives
with central charge. In the central basis, 
$\bD^{\pm \pm}$ coincide with $\pa^{\pm \pm}$,
while $\bD_\a^i$ and  $\bar{\bD}_{\ad i}$ are
\be
\bD^i_\a ~ = ~ \frac{\pa}{ \pa \q^\a_i } 
+ {\rm i} \, (\s^m \bar{\q}^i ) \pa_m 
- {\rm i}  \, \q_\a^i \, \D ~,\qquad 
\bar{\bD}_{\ad i} ~ = ~ - 
\frac{\pa}{\pa \bar{\q}^{\ad i} }
- {\rm i}\,  ( \q_i \s^m )_\ad \pa_m 
- {\rm i}  \, \bar{\q}_{\ad i} \, \D ~.
\ee
In the analytic basis which we mainly use, 
$\bD^+_{\hat{\a}}$ coincide with $D^+_{\hat{\a}}$
and the other derivatives are \cite{gikos}: 
$ \bD^-_{\hat{\a}} = D^-_{\hat{\a}} -2{\rm i}\, 
\q^-_{\hat{\a}}  \, \D$,  
$~\bD^{\pm \pm} = D^{\pm \pm} +{\rm i}\,\big(
(\q^\pm)^2 - (\bar{\q}^\pm)^2 \big) \D$.

The above global restriction on $\cV^{++}_5$
gets automatically accounted for if,
instead of using the representations
(\ref{param}) and (\ref{par2}), we start
decomposing the harmonic covariant 
derivative and gauge parameters with 
respect to the flat covariant derivatives
with central charge 
\bea
\cD^{++}_\D  &=&\bD^{++} + H^{++M} \bD_M +H^{(+4)} \bD^{--}
+ H^{+\hat{\a}} D^+_{\hat{\a}} + V^{++}_5 \D 
\label{paramcc} \\
\l  + \l_5 \D &=& \L^M \bD_M + \L^{++} \bD^{--} 
+ \L_5 \D  \;.
\label{parcc}
\eea
Then, the flat superspace limit would 
correspond to $V^{++}_5 = 0$. 
However, such a representation is
sensible only if the matter multiplets $U$ under
consideration are characterized by 
a constant central charge, 
$\D U^I = {\rm i} {M^I}_J U^J$,
with $M = ({M^I}_J ) $ a constant mass matrix
independent on the supergravity prepotentials. 
Such a situation appears,
for examples, for hypermultiplets described
by unconstrained analytic superfields.
However, it is well known that there 
exist $\cN=2$ supermultiplets which contain
finitely many auxiliary fields and possess
an intrinsic central charge. This means
that setting the central charge to be constant 
is equivalent to putting the theory on shell
(for example, this applies to the hypermultiplet
with $8 +8$ off-shell degrees of freedom).
To have a finite number of component fields
in such theories, one has to impose special
constraints on `primary' superfields $U$
in order that the series $\{ U, \D U, \D \D U, \dots\}$
containes only few functionally independent 
representatives. The constraints imposed 
must determine not only the field content 
but also specify the off-shell central charge
as a nontrivial functional of the supergravity 
prepotentials. For such theories, 
representation (\ref{paramcc}) is useless,
because the flat derivatives $\bD^{\pm \pm}$
and $\bD_M$ involve the `curved' central 
charge. 
 
In  representation (\ref{paramcc}) the requirement
$$
D^+_\a \cD^{++}_\D \F^{(p)} = 0
$$
for any analytic superfield $\F^{(p)}$, implies that 
the set of equations (\ref{prepconstr1}) should be 
extended to include one more relation
\be
D^+_\a V^{++}_5 - 2{\rm i} H^{+++}_\a = 0\;.
\label{prepconstr2}
\ee
Now, the general solution of the constraints 
(\ref{prepconstr1}) and (\ref{prepconstr2})
is given by eq.  (\ref{prepot1}) along with 
\be
V^{++}_5 =  \frac{{\rm i}}{4} (D^+)^2 G
- \frac{{\rm i}}{4} ({\bar D^+})^2 G
 + v^{++}_5 ~, \qquad \quad D^+_{\hat{\a}} 
v^{++}_5 = 0\;.
\label{prepot3}
\ee
The pre-gauge invariance (\ref{pre-gauge}) turns into
\bea
\d G & = & \frac{1}{4} (D^+)^2 \O^{--} + \frac{1}{4}
({\bar D}^+)^2 \breve{\O}^{--} ~, \non \\
\d v^{++}_5 &=& {\rm i} (D^+)^4 \O^{--}
- {\rm i} (D^+)^4 \breve{ \O}^{--} \;.
\label{pre-gauge2}
\eea
We see that the minimal multiplet is described by 
the two prepotentials $G$ and $v^{++}_5$,
the latter being a real analytic superfield.

Operator (\ref{parcc}) must move every analytic
superfield into an analytic one. This restricts 
the parameters $\L^M$ and $\L^{++}$ to have 
the form (\ref{prepot2}), while $\L_5$ reads
\be
\L_5 =  \frac{{\rm i}}{4} (D^+)^2 l^{--}
- \frac{{\rm i}}{4} ({\bar D}^+)^2 l^{--}
  + \hat{\l}_5 ~,
\qquad \quad D^+_{\hat{\a}}  \hat{\l}_5 = 0
\ee
with $\hat{\l}_5$ being an arbitrary 
real analytic superfield.

In the gauge (\ref{goodgauge}), $H^{++M}$, 
$H^{(+4)}$ and $V^{++}_5$ transform
as follows 
\bea 
& \d H^{++M} =\l \; H^{++M} - \cD^{++} \L^M ~,
\qquad \quad  
\d H^{(+4)} =\l \; H^{(+4)} - \cD^{++} \L^{++}  \non \\
& 
\d V^{++_5} =\l \; V^{++}_5 - \cD^{++} \L_5 
\label{trlawfinal}
\eea  
and hence
\be
\d v^{++}_5 =\l \; v^{++}_5 - \cD^{++} \hat{\l}_5 \;.
\label{v5trlaw}
\ee
As concerns the prepotential $G$,
from (\ref{trlawfinal}) we again deduce its 
transformation (\ref{Gtrlaw}).

\section{Supercurrent and multiplet of anomalies}
\setcounter{footnote}{0}
Given a matter system coupled to the minimal supergravity
multiplet, we define the supercurrent and multiplet of anomalies
\be 
\cJ = \frac{\d S}{\d G} ~, \qquad \quad 
\cT^{++} = \frac{\d S}{\d v^{++}_5}
\ee
with $S$ being the matter action. Here the variational derivatives 
with respect to the supergravity prepotentials 
are defined as follows
\be
\d S = \int {\rm d}^{12}z\, {\rm d}u \; \d G \;\frac{\d S}{\d G}
~~+~~ \int {\rm d}u \, {\rm d} \z^{(-4)}\; \d v^{++}_5 \; 
\frac{\d S}{\d v^{++}_5}\;.
\ee
As is seen, the supercurrent $\cJ$ is a real harmonic superfield,
$\breve{\cJ} = \cJ$, while the multiplet  of anomalies
 $\cT^{++}$ is a real analytic superfield,
$\breve{\cT}^{++} = \cT^{++}$, $D^+_{\hat{\a}} \cT^{++} = 0$. 
By construction,
both $\cJ$ and $\cT^{++}$ are inert with respect to the
central charge transformations.

The action is required to be invariant under pre-gauge transformations
(\ref{pre-gauge2}).  This means
\bea
\d S &=&  \frac{1}{4} \int {\rm d}^{12}z\, {\rm d}u \; \cJ\; (D^+)^2 \O^{--}
+{\rm i} \int {\rm d}u \, {\rm d} \z^{(-4)} 
\cT^{++} \;(D^+)^4 \O^{--} ~+~{\rm c.c.}\non \\
&=& \int {\rm d}^{12}z\, {\rm d}u \; \O^{--}
\left\{ \frac{1}{4} (D^+)^2 \cJ + {\rm i}\, \cT^{++} \right\}
~+~{\rm c.c.} = 0 \non
\eea
for arbitrary $\O^{--}$. As a consequence, we get
\be
\frac{1}{4} (D^+)^2 \cJ + {\rm i}\, \cT^{++} ~=~0 ~,
\qquad \quad
\frac{1}{4} ({\bar D}^+)^2 \cJ  -{\rm i}\, \cT^{++} ~=~0 
\;.
\label{conslaw2}
\ee

The action must be also invariant under the superspace 
general coordinate transformation group. The group 
acts on the prepotentials $G$ and $v^{++}_5$ 
according to eqs.  (\ref{Gtrlaw}) and (\ref{v5trlaw})
respectively. These transformations should be supplemented 
by those of the matter superfields. On shell, the invariance 
of $S$ with respect to (\ref{Gtrlaw}) and (\ref{v5trlaw}) turns out
to imply very strong restrictions on $\cJ$ and $\cT^{++}$, 
in addition to the conservation law  (\ref{conslaw2}).
Let us describe here the implications of general coordinate
invariance for the simplest and most interesting case 
of flat superspace when $G=v^{++}_5 =0$ (in general, the analysis
is basically the same but requires more involved technical tools). 
For such a background 
eqs. (\ref{Gtrlaw}) and (\ref{v5trlaw}) reduce to the 
linearized transformations 
\be 
\d G = - D^{++} l^{--} ~, \qquad \quad 
\d v^{++}_5 = - D^{++} \hat{\l}_5 \;.
\ee
Now, the invariance of $S$ with respect to the $l^{--}$ transformations 
means
\be
\d S =  - \int {\rm d}^{12}z\, {\rm d}u \; 
(D^{++} l^{--}) \,\cJ = 
\int {\rm d}^{12}z\, {\rm d}u \; l^{--} D^{++} \cJ = 0
\non
\ee
for arbitrary $l^{--}$, and hence 
\be 
D^{++} \cJ = 0\;.
\label{aa}
\ee
We see that the matter supercurrent in Minkowski superspace 
is $u$--independent, $J = J(z)$.
On the same ground, the invariance of $S$ with respect to 
the  central charge $\hat{\l}_5$--transformations implies
\be
D^{++} \cT^{++} = 0\;.
\label{bb}
\ee
The general solution of this equation in the central frame reads
\be
\cT^{++} (z,u) = \cT^{(ij)}(z) u^+_i u^+_j \;.
\ee
Since $\cT^{++} (z,u)$ has to be analytic, 
the multiplet of anomalies  $\cT^{(ij)}(z)$
satisfies eq. (\ref{trace}).

\section{Matter models in supergravity background}
\setcounter{footnote}{0}
In this section we will describe $\cN=2$ supersymmetric models, 
both with intrinsic central charge and models with constant central charge. 

\subsection{Models with intrinsic central charge}

We will use
analytic densities $\F^{(p)}_{ \{w\} }$
transforming as
\be 
\d \F^{(p)}_{ \{  w \} } = (\l + \l_5 \D )
\F^{(p)}_{ \{ w \} } + w\, \L \,
\F^{(p)}_{ \{ w \} } ~, \qquad \quad 
D^+_{\hat{\a}} \F^{(p)}_{ \{ w \} } =
D^+_{\hat{\a}} \d \F^{(p)}_{ \{ w \} } = 0
\ee
where the variation of the analytic subspace measure
${\rm d}u\, {\rm d} \z^{(-4)}$ with respect to general 
coordinate transformations (\ref{gctr})
is given by the analytic superfield
\be
\L \equiv (-1)^M D_M \L^M + D^{--} \L^{++}~,
\qquad \quad D^+_{\hat{\a}} \L = 0 \,.
\ee
We will be mainly interested in analytic densities 
$\J^{(p)}_{ \{p/2 \} }$ on which we can consistently impose the 
constraint
\be
\left( \cD^{++} + \cV^{++}_5 \D 
+ \hf p \,\G^{++} \right)
\J^{(p)}_{ \{p/2 \} } =  0
\label{basicconstraint}
\ee
where the analytic connection $\G^{++}$
is defined by \cite{gio}
\be
\G^{++} = (-1)^M   D_M H^{++M} 
+ D^{--} H^{(+4)} ~,
\qquad \quad D^+_{\hat{\a} } \G^{++} = 0 
\ee
and transforms as
\be
\d \G^{++} = \l \G^{++}
- \cD^{++} \L - 2 \L^{++} \;.
\ee
The above constraint turns
out to be gauge covariant only if
$p=2w $.

To construct a supersymmetric action,
let us specify an analytic density
$\J^{(2)}_{ \{ 1 \} }\equiv\cL^{++}$
subject to the constraint (\ref{basicconstraint}).
Then, the integral 
\be
S = \int {\rm d}u\, {\rm d} \z^{(-4)}\,
\cV^{++}_5 \, \cL^{++}
\label{actioncc}
\ee
proves to be invariant under the 
supergravity gauge transformations.
Indeed, since $\cL^{++}$ is an analytic
density of weight 1, and $\cV^{++}_5$
is a scalar superfield,
their product transforms into a  total derivative
\bea
\d \left( \cV^{++}_5 \, \cL^{++} \right) &=&
(-1)^M D_M \left( \L^M\, \cV^{++}_5 \, \cL^{++} \right)
+ D^{--} \left( \L^{++} \, \cV^{++}_5 \, \cL^{++} \right) ~, \non \\
&=& (D^+)^4 D^{--} \left( l^{--}
\cV^{++}_5 \, \cL^{++} \right) 
\eea
under (\ref{gctr}), and the action
(\ref{actioncc}) remains invariant.
Here we have used eq. (\ref{prepot2}).
As concerns the central charge transformations, we have 
\be
\d \cV^{++}_5 = - \cD^{++}\l_5 \qquad \quad
\d \cL^{++} = \l_5 \D \cL^{++}
\ee
and, modulo total derivatives, the variation of $S$ vanishes
\be
\d S = \int {\rm d}u\, {\rm d} \z^{(-4)}\,
\l_5 \left( \cD^{++} + \cV_5^{++} \D + \G^{++}
\right) \cL^{++} =0
\ee
as a consequence of \ref{basicconstraint}. 
The above prescription to construct 
supersymmetric invariants is a natural generalization 
of the action rule given in \cite{dikst} for $\cN=2$ 
rigid supersymmetric theories with gauged central charge.

Now, let us turn to a hypermultiplet with intrinsic
central charge in conformal supergravity background.
The hypermultiplet is described by a constrained analytic
superfield ${\bf q}^+ \equiv   \J^{(1)}_{ \{ 1/2 \} }$ and its
conjugate $\breve{{\bf q}}^+$. It can be shown that 
the analyticity of ${\bf q}^+$ and the 
basic  constraint
\be
\left( \cD^{++} + \cV^{++}_5 \D 
+ \hf  \,\G^{++} \right) {\bf q}^+ =  0
\label{a1}
\ee
determine the central charge $\D$ as a nontrivial 
operator depending on the supergravity prepotentials.
The hypermultiplet dynamics is described by the Lagrangian
\be
\cL^{++} = \hf \,
\breve{{\bf q}}^+ 
\stackrel{\longleftrightarrow}{ \D} 
{\bf q}^+ 
-{\rm i}\, m\, \breve{{\bf q}}^+ {\bf q}^+\;.
\label{c1}
\ee 
The corresponding equation of motion enforces
the central charge to be constant \cite{bs}
\be 
\frac{\d S}{\d q^+} = 0 \qquad \Longrightarrow
\qquad
\D {\bf q}^+ = {\rm i} \,m \,{\bf q}^+\;.
\label{b1}
\ee

\subsection{Models with constant central charge}

Let us consider a dual, for applications more useful 
description of the hypermultiplet in 
terms of an {\it unconstrained} analytic superfield
$q^+ (\z,u)$ and its conjugate $\breve{q}^+(\z,u)$.
The dynamical superfield is defined to transform
as a density of weight $1/2$
\be
\d q^+ = \left(\l + \l_5 \D \right) q^+
+ \hf \L q^+
\ee
and its central charge is chosen to be constant 
\be
\D q^+ = {\rm i}\,m\,q^+
\label{b2}
\ee
off shell. The dynamics is described in curved superspace
by the action
\be
S = - \int {\rm d}u\, {\rm d} \z^{(-4)}\,
\left\{ \hf \breve{q}^+ \stackrel{\longleftrightarrow}{ \cD}{}^{++}
q^+ + {\rm i} \, m \, \cV^{++}_5 \breve{q}^+ q^+ \right\}
\label{unconaction}
\ee
which reduces to that given in \cite{gios2} 
for $m=0$. The action is invariant 
under all local symmetries.
The corresponding equation of motion reads
\be 
\frac{\d S}{\d q^+} = 0 \qquad \Longrightarrow \qquad
\left( \cD^{++} + \cV^{++}_5 \D + \hf \G^{++}\right) q^+=0\;.
\label{a2}
\ee
Comparing eqs. (\ref{a1}) and (\ref{b1}) with (\ref{a2})
and (\ref{b2}), we see that the two hypermultiplet
models are equivalent.
But the equation of motion 
in one model turns into the off-shell constraint 
in the other and vice versa.

The basic advantage of this model 
is that off the mass shell the dynamical variable
$q^+$ is an unconstrained superfield independent on
the supergravity  prepotentials. That is why one can 
readily vary the action with respect to these prepotentials.
Using eqs. (\ref{prepot1}) and (\ref{prepot3})
gives
\be
\breve{q}^+ \cD^{++} q^+ +
{\rm i}\,m\, \cV^{++}_5 \breve{q}^+ q^+
= \breve{q}^+ \bD^{++} q^+ +
(D^+)^4 \Big\{ \breve{q}^+ G 
\bD^{--} q^+ \Big\} 
+ {\rm i}\,m\, v^{++}_5 \breve{q}^+ q^+\;.
\ee
We therefore obtain
\be
\cJ = - \hf \breve{q}^+ 
 \stackrel{\longleftrightarrow}{ \bD }{}^{--} q^+ ~,
\qquad \quad 
\cT^{++}= -{\rm i}\,m\, \breve{q}^+ q^+ \;.
\label{hypsupercurrent}
\ee
 
Let us compute the supercurrent and multiplet
of anomalies (\ref{hypsupercurrent})
in flat superspace where the equation of
motion  (\ref{a2}) becomes
\be
\bD^{++} q^+ = 0\;.
\ee
In the central basis, $\bD^{\pm \pm}$
coincide with $\pa^{\pm \pm }$, 
and the on-shell superfields read
\be
q^+ = q^i (z) u^+_i ~, \qquad \quad
\breve{q}^+  = {\bar q}_i (z) u^{+i} ~, 
\qquad \quad {\bar q}_i = \overline{q^i}\;.
\ee
Now, eq. (\ref{hypsupercurrent}) leads to 
\be 
\cJ= - \hf \,{\bar q}_i \,q^i ~, \qquad \quad 
\cT^{++} = \cT^{ij}(z) u^+_i u^+_j  ~, \qquad \quad
\cT^{ij} = {\rm i}\, m\, {\bar q}^{(i} q^{j)} \;.
\ee
What we have derived is exactly
the $\cN=2$ supercurrent and multiplet
of anomalies found by Sohnius \cite{sohnius}.

The above consideration can be generalized 
to the case of a general renormalizable super 
Yang-Mills system in curved superspace
with action 
\be
S = \frac{1}{2g^2} \,{\rm tr} \int {\rm d}^4 x\,
{\rm d}^4 \Theta \, \cE \,\cW^2
- \int {\rm d}u\, {\rm d} \z^{(-4)}\, \breve{q}^+ 
 \left\{ \hf  \stackrel{\longleftrightarrow}{ \cD}{}^{++}
+ {\rm i} \cV^{++}  + {\rm i}  \, \cV^{++}_5 M \right\} q^+\;.
\label{ymaction}
\ee
Here $\cV^{++} = \cV^{++}_I (\z, u) R_I$ is the Yang-Mills gauge 
superfield, and $\cW$ the corresponding covariantly chiral 
strength; $\cE$ denotes the $\cN=2$ chiral density \cite{muller,muller2}.
The constant mass matrix $M$ is required to be hermitian and to commute
with the gauge group, $[\cV^{++}, M] =0$.
In flat superspace, the corresponding supercurrent and multiplet 
of anomalies read
\be
\cJ = \frac{1}{g^2} \,{\rm tr} \left( {\bar \cW} \cW \right)
-\hf \breve{q}^+   \stackrel{\longleftrightarrow}{\nabla}{}^{--}
q^+ ~, \qquad \quad
\cT^{++}= -{\rm i}\, \breve{q}^+ M q^+ 
\label{ymsc}
\ee
where $\nabla^{--}$ denotes
the proper gauge covariant harmonic derivative. In the central 
basis, $\nabla^{--}$
coincides with $\pa^{--}$, and on shell 
\be 
q^+ = q^i (z) u^+_i ~, \qquad \quad
\nabla^{(i}_\a q^{j)} = {\bar \nabla}^{(i}_\ad q^{j)} = 0
\ee
where $\nabla^i_\a$ and ${\bar \nabla}^\ad_i$
denote ordinary $u$-independent 
gauge covariant derivatives. Therefore, from eq. 
(\ref{ymsc}) we get 
\be
\cJ = \frac{1}{g^2} \,{\rm tr} \left( {\bar \cW} \cW \right)
-\hf \, \bar q_i q^i ~, \qquad \quad
\cT^{ij}  = {\rm i}\,  {\bar q}^{(i} M q^{j)} \;.
\ee
It  is worth noting that (\ref{ymaction})
describes a curved superspace 
extension of the $\cN=4$ super Yang-Mills theory 
if $M=0$ and if $q^+$ transforms 
in the adjoint representation of the gauge group.

It is well known that $\cN=2$ Poincar\'e or de Sitter supergravity
cannot be formulated solely in terms of the minimal multiplet
\cite{dwsugra,bs}. To get a consistent action for Poincar\'e 
supergravity, one has to couple the minimal multiplet to an 
auxiliary multiplet whose role is to compensate 
%(all or part)
%local SU(2) transformations. 
some local transformations. 
Such a compensator may contain
finitely many \cite{dwsugra} or an infinite number \cite{gios2} 
of off-shell component fields. 
The three known minimal formulations \cite{dwsugra}
comprising $40+40$ off-shell degrees of freedom and their 
compensators are: (I) nonlinear multiplet; 
(II) hypermultiplet with intrinsic central charge
(\ref{a1}); (III) improved tensor multiplet. 
In principle, one can elaborate on non-minimal supergravity 
formulations with $n+n$ off-shell degrees of freedom, 
$40 < n < \infty$. Finally, there exists the maximal  
formulation \cite{gios2} whose compensator is a single
$q^+$ hypermultiplet considered in this subsection.
In all cases, the supergravity action is a sum of the action of the 
minimal multiplet and that for the compensator
\cite{dwsugra,gios2}.
%For example, its role 
%can play a  hypermultiplet with intrinsic central charge
%(\ref{a1}) under mild global assumptions
%; its scalar fields of lowest dimension should possess 
%nonvanishing values 
%\cite{dwsugra}. Then, the supergravity action 
%is given by a sum of the minimal multiplet action \cite{dwsugra,gios2}
%and (minus) the action (\ref{actioncc}, \ref{c1}); the choice $m=0$
%corresponds to Poincar\'e supergravity while $m \neq 0$
%to de Sitter supergravity \cite{dwsugra}. 

No matter what compensator we choose, 
it does not enter the minimal classical action 
(\ref{ymaction}) corresponding to general $\cN=2$ renormalizable 
SYM models. Therefore, the choice of compensator has no impact on the 
structure of the supercurrent at the classical level. 
The main effect of the compensator is to assure self-consistency
of the dynamics of the full supergravity-matter system.

If we give up the requirement of renormalizability, the compensator 
can tangle with $\cN=2$ matter. This is the case for general quaternionic 
off-shell sigma models in curved harmonic superspace
\cite{bgio}. But then we  deal with effective field theories
(e.g. low energy string actions)
and can treat the compensator as part of the matter sector
coupled to $\cN=2$ conformal supergravity.

As an example of more general dynamics, let us consider the 
$\cN=2$ rigid supersrymmetric
sigma model
\be
S = - \frac{1}{2} \, 
\int {\rm d}u\, {\rm d} \z^{(-4)}\, \left \{ \breve{q}^+ 
  \stackrel{\longleftrightarrow}{ \bD}{}^{++} q^+
+\frac{\l}{2}\, (  \breve{q}^+ q^+)^2 \right\}
\ee
with $\l$ the coupling constant.
The bosonic sector of this model describes the Taub--NUT 
gravitational instanton
with a scalar potential generated by the central charge.
To lift the model to curved superspace, one has to couple 
the dynamical superfields not only to the minimal 
supergravity multiplet, but also to 
an unconstrained analytic density $\o$ \cite{gios1,gios2}.
As a result, the coupling to supergravity is characterized
by $\cJ$ and $\cT^{++}$ given, in the flat superspace limit, 
by (\ref{hypsupercurrent})
along with the analytic superfield $\cT^{(+4)} = \d S /  \d \o
= - \frac{\l}{2} (\breve{q}^+ q^+ )^2$. The conservation
equations (\ref{conslaw2}) and (\ref{bb}) remain unchanged,
but eq. (\ref{aa}) gets modified to 
\be
D^{++} \cJ ~+~ D^{--} \cT^{(+4)} ~=~ 0
\ee 
and therefore $\cJ$ becomes $u$-dependent (note
that $(D^{++})^2 \cJ =0$, since $\bD^{++} (\breve{q}^+ q^+ )=
D^{++} (\breve{q}^+ q^+ )=0$ 
on shell).

%\footnote{
%It should be 
%stressed, however, that the compensator  given  in \cite{gios2} 
%enters, in general, the off-shell quaternionic 
%sigma models 
%coupled to $N=2$ supergravity  
%described  in \cite{bgio}. 
%According to \cite{gios2}, it is impossible to construct 
%an analytic superfield density by making use of the minimal 
%multiplet only, unlike the existence of the unique chiral 
%density \cite{muller}. 
%But such an analytic density is necessary to formulate 
%general off-shell matter couplings in supergravity background
%\cite{gios2,bgio}. 

\end{appendix}

\end{document}